\documentclass[a4paper,fleqn,usenatbib]{mnras}

\usepackage{amsmath}
\usepackage{amsfonts}
\usepackage{amsbsy}
\usepackage{amssymb}
\usepackage{amsbsy}

\usepackage[T1]{fontenc}
\usepackage{ae,aecompl}

\usepackage{graphicx}
\usepackage{gensymb}
\usepackage{float}
\usepackage[utf8]{inputenc}
\usepackage{tensor}
\usepackage{bm}
\usepackage{enumerate}

\title[]{Magnetorotational instability and dynamo action in gravitoturbulent astrophysical discs}

\author[]{
A. Riols,$^{1}$ H. Latter $^{1}$
\\
$^{1}$Department of Applied Mathematics and Theoretical Physics, University of Cambridge, Centre for Mathematical Sciences, \\
Wilberforce Road, Cambridge CB3 0WA, UK.}

\date{Accepted XXX. Received YYY; in original form ZZZ}

\pubyear{2017}

\begin{document}
\label{firstpage}
\pagerange{\pageref{firstpage}--\pageref{lastpage}}
\maketitle

\begin{abstract}
Though usually treated in isolation, the magnetorotational and
gravitational instabilities (MRI and GI) may coincide 
at certain radii and evolutionary stages of protoplanetary discs
and active galactic nuclei. 
Their mutual interactions
could profoundly influence several important processes, such as
accretion variability and outbursts, fragmentation and disc
truncation, or large-scale magnetic field production. 
Direct numerical simulations of both instabilities are computationally
challenging and remain relatively unexplored. In this paper, we aim
to redress this neglect
via a set of 3D
vertically stratified shearing-box simulations, combining self-gravity
and magnetic fields.
We show that gravito-turbulence
greatly weakens the zero-net-flux MRI. 
In the limit of efficient cooling (and thus enhanced GI), the MRI is
completely suppressed, and yet
strong magnetic fields are sustained by the gravitoturbulence. 
This turbulent `spiral wave' dynamo may
have widespread application, especially in galactic discs. 
Finally, we present preliminary work showing that a strong
net-vertical-flux revives the MRI and supports a magnetically
dominated state, in which the GI is secondary.
\end{abstract}




\section{Introduction}

The magnetorotational and
gravitational instabilities (MRI and GI) are perhaps the most efficient and
commonly invoked mechanisms driving angular momentum transport in
astrophysical accretion disks \citep{balbus99,armitage11}. 
In order to work, the MRI requires the gas to be sufficiently coupled
to any latent magnetic field, a condition that can be framed in terms
of Elsasser numbers for Ohmic and ambipolar diffusion, in addition to the Hall
effect \citep[e.g.][]{wardle99, balbus01,kunz04}. Ascertaining the 
prevalence of the MRI in protoplanetary (PP) discs,
especially, is a vexed business involving poorly known inputs
such as the amount and size of dust, the nature of the ionising radiation
field, and disc geometry. The onset
of GI is more straightforward, developing in sufficiently 
massive or thin discs. The criterion for 
axisymmetric GI is:
\begin{equation}
Q=\dfrac{c_s\Omega}{\pi G \Sigma_0}\lesssim 1
\label{mass_eq}
\end{equation} 
where $Q$ is the Toomre parameter, $c_s$ is the sound speed, $\Omega$
the orbital frequency, and $\Sigma_0$ the background surface density
\citep{toomre64}. Nonaxisymmetric GI occurs for $Q$s somewhat larger, 
leading to a ``gravito-turbulent" state, in the case of inefficient
radiative cooling, 
or fragmentation, in the opposite case.  
\\

A fundamental and unresolved problem concerns the interaction
between these two instabilities. 
In AGN discs, both are thought to be excited:
GI beyond some $10^3$ gravitational radii, and the MRI within some
second critical radius. In sufficiently luminous sources ($\dot{M} \gtrsim 10^{-2} M_\odot$
yr$^{-1}$), 
the two instability regions can, in fact, overlap \citep{menou01}, and
this is certainly the case if the disc intercepts a reasonable fraction of
the central X-ray source's emission, as it might do if warped
\citep[e.g.  NGC 4258,][]{neufeld95}. The mutual interaction 
of the two instabilities may impact, in particular, on the AGN disk
truncation problem --- especially if the MRI softens or
suppresses GI. It may also be a source of the rich accretion
variability exhibited by these sources \citep[][and see below]{menou01,peterson01}. 
\\

Young PP discs are relatively massive compared to the
central accreting protostar and so usually undergo some period of GI
early in their lifetime
\citep{kratter16};
for instance, roughly 50\% of class 0 and 10-20\% of 
class I sources might possess unstable outer
radii 
\citep{tobin13, mann15}. 
Meanwhile, the gas is subject to an
ionising flux of cosmic rays and stellar X-rays that may be
sufficient to couple the gas to the magnetic field of the collapsing
cloud and hence permit the emergence of the MRI in some
form. While the conditions for the onset of MRI are not especially favourable in
the later epochs of PP disc lifetimes \citep[e.g.][]{lesur14}, the shorter class 0 stage
may offer a more amenable environment. 
The importance of the MRI/GI interaction here lies in its potential
role in suppressing (or enhancing) disk
fragmentation and, consequently, the formation of long-period exoplanets \citep{rice15}.
\\

The GI and MRI synergy may also bear on the striking accretion
outbursts that slightly older PP disks endure, as emulated by the
FU-Orionis and EX-Lupi variables \citep{hartman96,evans09,aguilar12}.
A popular model for these events invokes a 
`gravo-magneto' limit cycle, according to which (a) material piles up
in a dead zone, where neither GI nor MRI is active, until (b) a sufficiently large mass 
initiates GI, which in turn (c) thermally ionizes the gas
and instigates MRI, and then finally (d) the excess mass is accreted by the
MRI in an eruptive event \citep{armitage01,zhu10,martin11,martin12}.
Low luminosity AGN, which may also support ``dead radii'', could
undergo similar dynamics \citep{menou01}.
 \\

In this paper we put aside the issue of outbursts and `dead zones' and
focus on the more tractable and fundamental question of the GI's and
MRI's coexistence: 
can the disk sustain a turbulent quasi-steady
state issuing from, and blending, both instabilities? We would like
to describe the
properties of such a state (if it exists), assess how it depends on the
cooling efficiency of the gas,
and determine whether the MRI suppresses or accelerates fragmentation.  
Note that this problem was first tackled  by \citet{fromang04} and
\citet{fromang2005} some 15 years ago in pioneering global simulations. 
They showed that the transport
properties of both instabilities are not additive: MHD turbulence can
reduce the rate of GI angular
momentum. Most of the simulations were poorly
resolved (5-10 points per scale height), imposed magnetic fields of almost thermal strengths, 
and were run for only a few orbits. One aim of
this paper is to
improve on these points and survey a larger region of parameter space.
\\

In addition, we are interested in the coupling between 3D
gravito-turbulence
and magnetic fields generally, and in the potential of gravito-turbulence to sustain
a large or small-scale dynamo in the absence of the MRI. This problem
could be relevant to situations where the MRI is quenched by GI 
or non-ideal MHD, as certainly might be the case in the poorly
ionised disks of interest. Preliminary analyses have been carried out in 2D
\citep{kim2001,riols16a} and
 3D SPH global simulations \citep{forgan17}, but in both cases
 the numerical methods were unsuitable for establishing
 the existence of a GI-driven dynamo.
\\

We performed 3D shearing
box simulations of gravito-turbulent discs with magnetic fields and
vertical stratification using the code PLUTO endowed with a simple linear
cooling law and characterised by a single cooling time $\tau_c$. Most of our simulations
focus on the 
zero-net-flux magnetic field configuration. The physical
problem is computationally challenging since the MRI
manifests on length-scales much shorter than the scaleheight $H$,
whereas GI
exhibits characteristic scales much longer than $H$. We managed a
spatial resolution of $\sim 26$ cells per $H$ in both
radial and azimuthal directions, in boxes of horizontal size $20H$, 
and ran simulations for $\sim$ 100
orbits.
Our Poisson solver captures 
the full gravitational potential (including small-scales) and has been
designed to deal with non-periodic vertical boundary conditions
\citep{riols17b}.
\\

 Our first main result is relatively simple: the zero-net-flux MRI is
difficult to sustain in the presence of gravitoturbulence. Its survival depends on the cooling
efficiency of the gas (in effect, a proxy for the strength of GI). 
For $\tau_c \gtrsim 100 \,\Omega^{-1}$, a sluggish MRI
persists at $z\simeq H$ and coexists with GI, but fails to produce
vigorous and regular butterfly dynamo patterns. This weakly magnetized state
is characterized by a mix of large-scale spirals and thin axisymmetric
structures. At shorter cooling times $\tau_c< 100\,\Omega^{-1}$, 
the MRI is completely quenched and the flow
dominated by gravitoturbulence. Nevertheless, the magnetic field
is amplified to
equipartition strengths by a dynamo process relying
exclusively on the GI spiral waves. This is our second main result. 
It is possible that this `spiral
wave dynamo' is relevant to magnetic field generation in galactic
disks, and is an attractive alternative to the ubiquitous mean-field
dynamo models used in the field. Fragmentation
occurs when $\tau_c
\lesssim 2\Omega^{-1}$,  a similar value to that computed in
hydrodynamics. 
Finally, we examine the case of a relatively strong
net-vertical magnetic flux, with a mid-plane beta of approximately 200.
The results are radically different to ealier, with the disk
supporting
a magnetically dominated flow similar to that witnessed by \citet{salvesen16},
 in which GI is secondary. Further net-flux results are reserved
for a separate paper. The key point is that there exists at least one regime
in which the MRI is dominant and GI suppressed.\\

The structure of the paper is as follows: first, in
Section \ref{sec_model}, we introduce the basic equations of the
problem and present our numerical setup and some useful
diagnostics. In Section \ref{sec_MRI}  we study the MRI turbulence in
the limit $Q\to\infty$ (no self-gravity), in order to obtain a reference
state to compare with simulations incorporating self-gravity. The aim of this
section is also to identify the properties of MRI turbulence in large
boxes and test their convergence on resolution. In Section
\ref{sec_GIMRI}, we present MHD turbulent runs with
GI and explore different cooling time regimes. In particular
we compare simulations initialized from different states (either pure
hydrodynamic GI or pure MRI with $Q\to\infty$).
 In Section
\ref{sec_dynamo}, we characterise the nature of the dynamo process,
and shows that GI can amplify magnetic fields efficiently and
independently of the MRI. Finally, in Section \ref{sec_discussion}, we discuss the
possible implications of our results 
for astrophysical discs.

\section{Numerical model}
\label{sec_model}
\subsection{{Governing equations}}

We adopt the shearing box \citep{goldreich65}, a local Cartesian model
of an accretion disc, 
because it offers potentially excellent resolution and hence permits
us to capture both the MRI small-scale turbulence and the GI
large-scale motions \citep[see complementary justifications in
][]{riols17b}.
 In this model, the differential rotation is approximated locally
by a linear shear flow and a uniform rotation,
$\boldsymbol{\Omega}=\Omega \, \mathbf{e}_z$. 
We denote by $(x,y,z)$
the radial, azimuthal, and vertical directions respectively,  
and refer to the $(x,z)$ projection of a vector field as its `poloidal'
component and its $y$ component as its `toroidal' one. 
The gas is ideal, its pressure $P$ and density $\rho$ related by $\gamma P=\rho c_s^2$, where $c_s$ is the
sound speed and $\gamma$ the ratio of specific heats. The pressure is
hence related to internal energy $U$ by $P=(\gamma-1)U$. 
We neglect molecular viscosity but consider a non-zero magnetic
diffusivity $\eta$ in some cases. 

The evolution of density ${\rho}$, total velocity $\mathbf{v}$,
magnetic field $\mathbf{B}$, 
and internal energy $U$ obeys
\begin{equation}
\dfrac{\partial \rho}{\partial t}+\nabla\cdot \left(\rho \mathbf{v}\right)=0,
\label{mass_eq}
\end{equation}
\begin{equation}
\frac{\partial{\mathbf{v}}}{\partial{t}}+\mathbf{v}\cdot\mathbf{\nabla
  v} +2\boldsymbol{\Omega}\times\mathbf{v} =-\nabla\Phi
  -\dfrac{\mathbf{\nabla}{P}}{\rho}+\dfrac{(\nabla\times \mathbf{B})\times\mathbf{B}}{\rho},
\label{ns_eq}
\end{equation}
\begin{equation}
\frac{\partial{\mathbf{B}}}{\partial{t}} =\nabla\times(\mathbf{v}\times\mathbf{B})+\eta\,\mathbf{\nabla}^2\mathbf{B},
\label{magnetic_eq} 
\end{equation}
\begin{equation}
\dfrac{\partial U}{\partial t}+\nabla\cdot (U\mathbf{v})
  =-P\nabla\cdot\mathbf{v}-\dfrac{U}{\tau_c},
\label{int_energy_eq}
\end{equation}
where the total velocity field can be decomposed into the background
 orbital shear and a perturbation $\mathbf{u}$: 
\begin{equation}
\mathbf{v}=-S x\,  \mathbf{e}_y+\mathbf{u},
\end{equation}
with
$S=(3/2)\,\Omega$ for a Keplerian equilibrium.
$\Phi$ is  the sum of the tidal potential in the local frame
$\Phi_c=\frac{1}{2}\Omega^2 z^2-\frac{3}{2}\Omega^2\,x^2$  and the
gravitational potential induced by the disc itself, $\Phi_s$. The
latter obeys the Poisson equation 
\begin{equation}
\mathbf{\nabla}^2\Phi_s = 4\pi G\rho.
\label{poisson_eq}
\end{equation}
We assume that the cooling in the internal energy equation
(\ref{int_energy_eq}) 
is a linear function of $U$ with a typical timescale $\tau_c$ referred to as the `cooling
time'. This prescription is not especially realistic but allows us to
control 
the rate of energy loss via a single parameter. We also neglect thermal conductivity. Finally, $\Omega^{-1}=1$ defines our
unit of time and $H_0=1$ our unit of length, where 
$H_0$ is the disc scale height, defined to be the
ratio $c_{s_0}/\Omega$, with
 $c_{s_0}$ denoting the sound speed in the midplane of a
 non-self-gravitating hydrostatic disc ($Q\rightarrow \infty$).

\subsection{Numerical methods}

The numerical methods are identical to those used by
\citet{riols17b}. 
The treatment of the 
boundary conditions for $\mathbf{B}$, absent in our hydrodynamical
study, 
is detailed in Section \ref{boundary_cond}.
\subsubsection{Code} 
\label{code}
We use the Godunov-based PLUTO code
\citep{mignone2007} to perform direct numerical simulations of the three-dimensional flow in the shearing box frame. The box has a finite domain of
size $(L_x,L_y,L_z)$, discretized on a mesh of $(N_X,N_Y,N_Z)$ grid
points. The numerical scheme uses a conservative finite-volume method
that solves the approximate Riemann problem at each inter-cell
boundary. It is well adapted to highly compressible flow and
reproduces the behaviour of conserved quantities like mass, momentum,
and total energy. The Riemann problem is handled by the HLLD solver,
suitable for MHD. An orbital advection algorithm is used to
increase the computational speed and better deal with the large background
shear flow. Finally, because PLUTO conserves the total energy, the
heat equation Eq.(\ref{energy_eq}) is not solved directly. The code,
consequently, captures the
irreversible heat produced by shocks due to numerical diffusion,
consistent with the
  Rankine Hugoniot conditions.
 The divergence of $\mathbf{B}$ is ensured to be 0 by the
 constrained-transport algorithm of PLUTO.

\subsubsection{Poisson solver}
\label{poisson_solver}
The 3D self-gravitating potential is computed as in \citet{riols17b}. For each plane of altitude
$z$, we compute $\hat{\rho}_{k_x,k_y} (z)$, the direct 2D Fourier transform of the density in a frame comoving with the shear, and solve a Helmholtz equation for the potential in Fourier space:
\begin{equation}
\left[\dfrac{d^2}{dz^2} -k^2\right] \hat{\Phi}_{k_x,k_y} (z) = 4\pi G \hat{\rho}_{k_x,k_y} (z),
\label{eq_helmhotz}
\end{equation} 
with $\hat{\Phi}_{k_x,k_y} (z)$ the planar Fourier transform of the
disc potential 
and $k=k_x^2+k_y^2$ the horizontal wavenumber. This equation is solved in the
complex plane by means of a 4th order finite difference scheme and a direct
inversion method (see \citet{riols17b} for more details). We then
compute the inverse Fourier transform of the potential and shift it
back to the initial frame. The gravitational forces are obtained by
computing the derivative of the potential in each direction.

 Unlike
methods based on 3D Fourier decomposition, which generally assume
periodic or vacuum boundary conditions for the potential
\citep{koyama10,shi2014}, our code can handle any kind of
boundary. The stratified disc equilibria, as well as the linear
stability of these equilibria, have been tested to ensure that our
implementation is correct (see appendices in \citet{riols17b}). Note that self-gravity is added as a source term in the momentum and energy equation, and not as a flux term as in the ATHENA code \citep{jiang13}.

\subsubsection{Boundary conditions}
\label{boundary_cond}

Boundary conditions are periodic in $y$ and shear-periodic in $x$
(Hawley et al.~1995). In the vertical direction,  we use a standard
outflow condition for the velocity field but enforce hydrostatic
balance in the ghost cells for pressure, taking into account the large
scale vertical component of  
self-gravity (averaged in $x$ and $y$). In this way we significantly  
reduce the excitation of waves near the boundary. See Appendix A of \citet{riols17b}. 

For the gravitational potential, we impose \begin{equation}
\dfrac{d}{dz} \Phi_{k_x,k_y} (\pm {L_z}/{2}) = \mp k\Phi_{k_x,k_y} (\pm {L_z}/{2}).
\end{equation} 
This condition is an approximation of the Poisson equation in the
limit of low density. In addition, we enforce  a density floor of $10^{-4}\, \Sigma/H_0$ which
prevents the timesteps getting too small due to evacuated regions near
the vertical boundaries. \\

For the magnetic field, we use the so-called ``vertical field" (VF)
boundary conditions where $B_x=B_y=0$ and $dB_z/dz=0$. They correspond
to a version of vacuum boundary conditions, appropriate for the
surface of a disc and have been used in a number of stratified MRI
simulations \citep{branden95,gressel10,oishi11,kapyla11}. The mean
horizontal magnetic field (or total flux) is not conserved and is
allowed 
to vary in the computational domain because of the boundary
conditions, 
even if the field initially has zero net $B_x$ and $B_y$.  
One concern with this set-up is that a mean-field dynamo might
be artificially sustained  by the VF conditions. Previous studies,
however, suggest that a large-scale magnetic field (modulated in the $z$
direction) is maintained even with periodic vertical boundary
conditions. This tells us that such a mean field 
is physical and not an artifact of
VF boundaries. Note that the presence of a mean field,
insensitive to diffusion, has 
obvious consequences for the existence and properties of dynamos in astrophysical regimes \citep{kapyla11,oishi11}. 
Finally, VF boundaries are known to produce spurious
currents near the boundaries, 
but this effect does not significantly affect the bulk properties of the
flow \citep{ziegler01,oishi11}.

\subsection{Simulation setup}

\subsubsection{Box size and resolution}

We expect the large-scale spiral waves excited by GI to possess a
radial lengthscale $\lambda \gtrsim H\,Q$ in the gravito-turbulent
regime. In order to capture these waves, while affording reasonable
resolution and numerical feasibility, we employ a box of intermediate
size $L_x=L_y=20 \, H_0$. The vertical domain of the box
spans $-3\,H_0$ and $3\,H_0$. 

Numerical resolution is chosen in such a
way that the most unstable non-axisymmetric MRI modes are resolved by
more than 20 points. In most runs with MRI and GI, a resolution of
$26$ points per $H_0$ is used in the horizontal directions. The total
resolution is $512 \times 512 \times 128$. Note that the dependence of
MRI on resolution is addressed in Section \ref{convergence_MRI} with
some small box runs ($L_x=2\, H_0$). 

\subsubsection{Parameters and initial equilibria}
\label{ic}

For all simulations presented in this paper, we use a fixed heat
capacity ratio $\gamma=5/3$. 
If mass is lost through the
vertical boundaries, it is replenished near the midplane so that the
total mass in 
the box is maintained constant.  
We checked that the mass injected at each orbital period is negligible
compared to the total 
mass (less than 1\% per orbit), and does not affect the results. 
The small vertical outflow also removes
thermal energy from the box, which is not replenished, and hence supplies an
additional means of cooling.  

The net vertical magnetic flux
is conserved in the box; a simulation is designated
 ``zero net flux" when the box averaged $B_z$ is zero. 
In the case of pure MRI without self-gravity, the initial equilibrium
is the classical hydrostatic disc. For pure hydrodynamic GI
simulations, the density equilibrium depends on $z$ in a non-trivial
way. We solve for the nonlinear set of equations that describes a
polytropic and self-gravitating disc equilibrium \citep[see Section
2.2 of][]{riols17b}. In both cases, random non-axisymmetric density
and velocity perturbations of finite amplitude are injected at $t=0$
to initiate the turbulent state. In pure MRI runs we
added a large scale sinusoidal toroidal field (modulated in
$z$) to trigger the MRI. For hybrid simulations, mixing MHD and GI,
initial conditions are generally computed from a pre-existing
turbulent state (either pure GI or pure MRI) and will be
specified in the corresponding sections. 

\subsection{Diagnostics}
\label{alpha}
\subsubsection{Averages}
To analyse the statistical behaviour of the turbulent flow, 
we define two different volume averages of a quantity $X$. The first is the standard average:
\begin{equation}
\left<X \right>=\frac{1}{L_xL_yL_z}\int_V X\,\, dV, 
\end{equation}
where $V$ denotes the volume of the box. The second is the density-weighted average:
\begin{equation}
\left<X \right>_w=\dfrac{\int_V \rho X\,\, dV}{\int_V \rho \,\, dV}.
\end{equation}
We also define the horizontally averaged vertical profile of a dependent variable: 
\begin{equation}
\overline{X}(z)=\dfrac{1}{L_xL_y} \int\int X\,\, dxdy.
\end{equation}

An important quantity that characterizes self-gravitating discs 
is the average 2D Toomre parameter defined by
\begin{equation}
Q=\dfrac{\left\langle c_s\right\rangle_w \Omega}{\pi G  \Sigma },
\end{equation}
where  $\Sigma=L_z \left<\rho \right>$ is the disc's mean
surface density.

Another useful quantity is the
coefficient $\alpha$ which measures the turbulent angular momentum
transport. 
This quantity is the total stress
(summing the gravitational $G_{xy}$, Reynolds $H_{xy}$,
and Maxwell stresses $M_{xy}$) divided by the average pressure:
\begin{align}
\label{def_alpha}
\alpha=\dfrac{\left\langle
  H_{xy}+G_{xy}+M_{xy} \right\rangle}{\langle P\rangle},
\end{align} 
where
\begin{align*}
H_{xy}=\rho u_xu_y,
\quad G_{xy}=\dfrac{1}{4\pi G}\dfrac{\partial \Phi}{\partial
  x}\dfrac{\partial \Phi}{\partial y} \quad \text{and} \quad M_{xy}=-B_xB_y.
\end{align*}

\subsubsection{Total energy budget}

In order to determine the energy budget, we introduce
the average kinetic, magnetic,
gravitational, and internal energies, denoted by
\begin{align*}
 E_c=\frac{1}{2}\langle\rho \mathbf{u}^2\rangle, \quad
E_m=\frac{1}{2}\langle\mathbf{B}^2\rangle, \quad E_G=~\langle \rho \Phi
\rangle,
\end{align*}
and $E_T={\langle U\rangle}=(\gamma-1)\langle P \rangle$,
respectively. 
In the shearing box with outflow vertical boundary conditions, the evolution of the averaged total energy $e$ follows: 
\begin{equation}
\label{energy_eq1}
\dfrac{\partial}{\partial t} \left\langle e+\frac{1}{8\pi G}\vert\mathbf{\nabla \Phi}\vert^2\right\rangle+\mathcal{F}_z[e+P]=\left(\alpha(\gamma-1)S-\dfrac{1}{\tau_c}\right) E_T,
\end{equation}
where
\begin{equation}
 e=\frac{1}{2}(\rho \mathbf{u}^2+\mathbf{B}^2)+\rho\Phi+(\gamma-1) P,
 \end{equation}
 and
\begin{equation}
\mathcal{F}_z[X]=\left[\int\int u_z X \,\,dx dy\right] ^{L_z/2}_{-L_z/2} 
\end{equation}
is the net vertical flux of a quantity across the vertical
boundaries. Equation (\ref{energy_eq1}) implies that the radial flux
of angular momentum is the only source of energy in the system that
can balance explicit cooling and the losses through the vertical
boundaries. 

Energy is extracted from the shear by the turbulent flow and is irremediably converted into heat. The flux term on the left hand side can be physically associated with a wind that removes energy from the disc. We define an appropriate ``wind cooling rate" as:
\begin{equation}
\tau_w(t)^{-1}=\dfrac{\mathcal{F}_z[e+P]}{E_T}.
\label{eq_tw}
\end{equation}
If turbulence is in a steady state then we expect
\begin{equation}
\label{energy_eq}
\alpha (t)(\gamma-1)S=\dfrac{1}{\tau_c}+\dfrac{1}{\tau_w(t)}.
\end{equation} 
This relation is very similar to \citet{Gammie2001} but includes
vertical losses through the boundaries. Note that replenishing mass in the
midplane and imposing a density threshold does not introduce any loss
or gain of total energy. It can alter the internal energy
budget but the changes are small and will always happen on a timescale $
\gtrsim(\Omega\alpha)^{-1}$. 

It is worth pointing out that, in the
case $\tau_c=\infty$ (no explicit cooling),  a thermodynamic equilibrium is
reached when the turbulence produces as much energy as it is able to
expel vertically. Then the transport efficiency $\alpha$ directly
depends on the wind properties and flux transport at the
boundary. This itself depends on various quantities, such as the disc
magnetization, the turbulent activity (related to $\alpha$) but also
the numerical details of the code, in particular the vertical box size
\citep{fromang13}. 

\subsubsection{Spectra and small/large-scale ratios}

To analyse the structure and size of turbulent eddies, it is most convenient to study the
flow in Fourier space.  We denote by  $\hat{\mathbf{u}}^e(k_x,k_y,z)$ and  $\hat{\mathbf{B}}^e(k_x,k_y,z)$ the horizontal 2D spectra (in
Eulerian wavenumbers) of the turbulent velocity and magnetic field, for a given
altitude $z$, and averaged over a given period of time $T$. 
These quantities are calculated via a method described by
\citet{riols17b} (Section 2.5.2), 
using the FFT algorithm. The 2D kinetic and magnetic energy power spectra are then defined as 
\begin{equation}
E_{K}(k_x,k_y,z)=\frac{1}{2} \Sigma H_0^{-1} \left\vert \hat{\mathbf{u}}^e(k_x,k_y,z)\right\vert^2,
\end{equation}
\begin{equation}
E_{M}(k_x,k_y,z)=\frac{1}{2} \left\vert\hat{\mathbf{B}}^e(k_x,k_y,z)\right\vert^2,
\end{equation}
We define next the time and $k_x$-averaged 1D spectrum 
\begin{equation}
\mathcal{E}_{K}(k_y,z)=\dfrac{1}{k_{x_0}}\int E_K(k_x,k_y,z)\,d k_x,
\end{equation}
where $k_{x_0}=2\pi/L_x$.
To quantify the importance of small-scale motions 
relative to large-scale motions, we introduce the $z$-dependent ratio:
\begin{equation}
\Lambda^{n,p}_{K}(z)= \dfrac{\mathcal{E}_K(n\pi H_0^{-1},z)}{\mathcal{E}_K(p\pi H_0^{-1},z)},
\end{equation}
with $n\gg p$. At a given height, this diagnostic tells us how much
kinetic energy is in long azimuthal scales ($=2H_0/n$) vis-a-vis short
scales ($=2H_0/p$).
Similarly we define $\mathcal{E}_{M}(k_y,z)$ and
$\Lambda^{n,p}_{M}(z)$ for the magnetic field.

\begin{figure}
\centering
\includegraphics[width=\columnwidth]{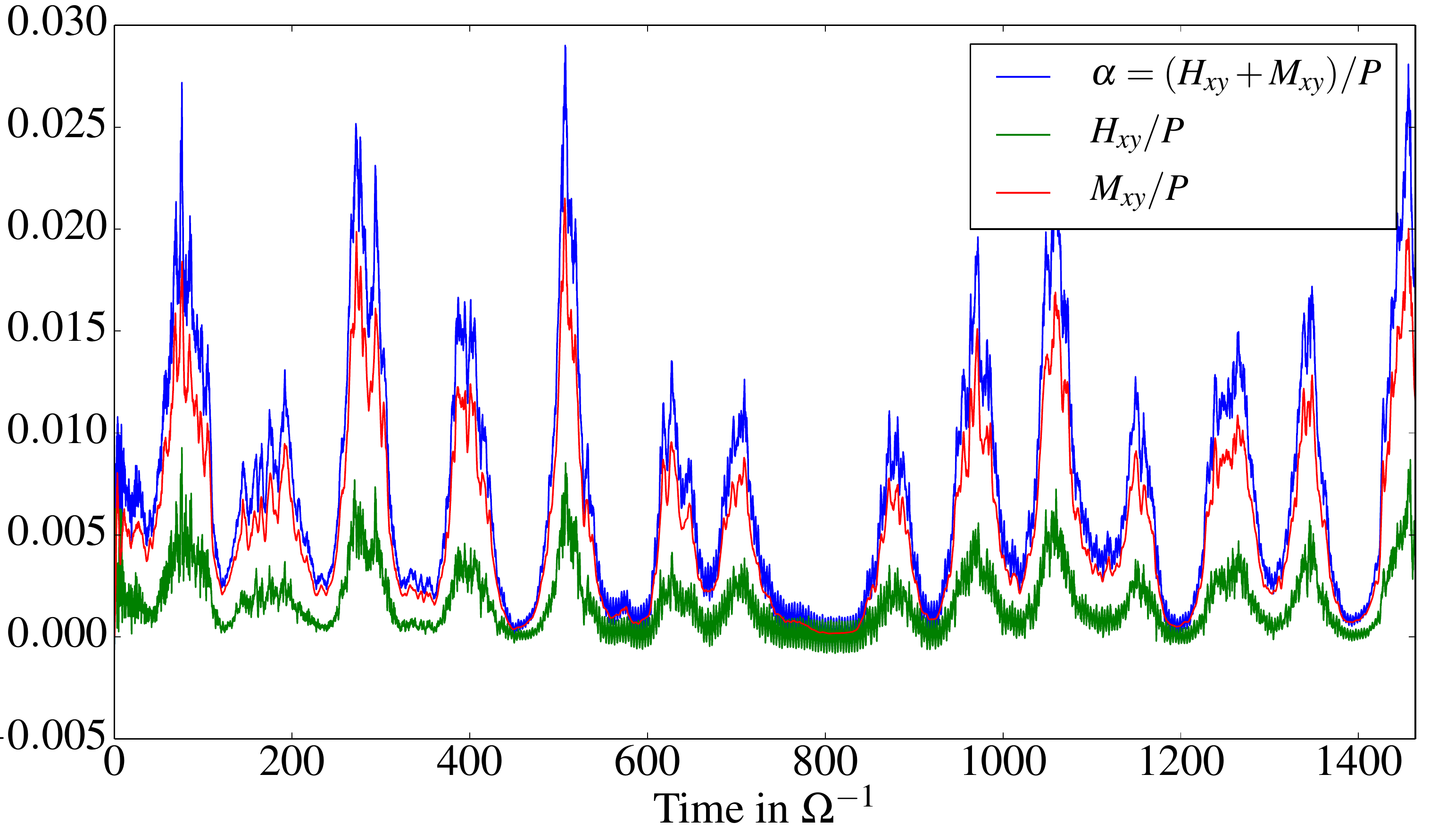}
 \caption{Time-evolution of the total stress (blue), Reynolds (green)
   and Maxwell (red) stresses, all normalized by mean pressure, 
for simulation MRI-S1 ($L_x=2\,H_0$, $L_y=4\,H_0$ and $L_z=6\,H_0$)}
\label{fig_mri_small}
 \end{figure}

\section{MHD simulations without self-gravity ($Q\rightarrow \infty$)}
\label{sec_MRI}

Before including self-gravity, it is necessary  to understand
the properties of MRI-driven turbulence in numerical models 
involving diabatic gas and very large horizontal domains.
While there is a rich literature describing the MRI in isothermal
stratified shearing boxes, only a few studies have dealt with the
thermodynamic
evolution of an ideal gas subject to non-adiabatic processes
\citep{turner03,hirose11,bodo12,gressel13,mcnally14,hirose14,coleman17}. Some
of these have highlighted the ability of MRI turbulence to
sustain convective motions that alter the system's turbulent transport and
dynamo properties.  However, none exactly correspond
to our setup. 

 In this section we perform our own
pure MRI simulations that can be directly compared to our later mixed GI/MRI
simulations. We stress that in this section we omit explicit cooling,
so that $\Lambda=0$ (i.e.\ $\tau_c\to\infty$), 
and thus the box is permitted to slowly heat up
via turbulent dissipation. Note that some of this energy will be lost
 due to weak outflows through the box's vertical
boundaries. The effective cooling timescale of these outflows ranges
between roughly $150\Omega^{-1}$ and $250\Omega^{-1}$. \\

\begin{table*}    
\centering 
\begin{tabular}{c c c c c c c c c c c}          
\hline                        
Run & Resolution & Time ($\Omega^{-1}$) & $\tau_c \,(\Omega^{-1})$ & $Q$ & $E_T$ & $E_c$ & $E_m$ & $H_{xy}$ & $G_{xy}$ & $M_{xy}$ \\   
\hline  
	MRI-S1	& $52\times 104 \times 128$ & 1500 & $\infty$ & $\infty$ &  0.62 & 0.0029 & 0.0096 & $7.5\times 10^{-4}$ & 0 & $0.00250$\\
	MRI-S2	& $64\times 128 \times 128$ & 1500 & $\infty$ & $\infty$ &  0.59 & 0.0029 & 0.0087 & $6.6\times 10^{-4}$ & 0 & $0.00215$\\
	MRI-S3-HD	& $128\times 256 \times 256$ & 1200 & $\infty$ & $\infty$ &  0.48 & 0.0020 & 0.0055 & $4.8\times 10^{-4}$ & 0 & 0.00175\\
	MRI-L1 	& $512^2\times 128$ & 1500 & $\infty$ & $\infty$ &  0.408 & 0.0027 & 0.0041 & $3.6\times 10^{-4}$ & 0 & $8\times 10^{-4}$\\    	
	\hline 
	MRISG-200 	& $512^2\times 128$ & 300 & $200$ & $1.50$ &  0.537 & 0.015 & 0.0016 & $0.0015$ & 0.0013 & $2.6\times 10^{-4}$ \\    
	MRISG-100 	& $512^2\times 128$ & 600 & $100$ & $1.43$ &  0.485 & 0.021 & 0.0017 & $0.0019$ & 0.0020 & $3.5\times 10^{-4}$ \\    
	MRISG-20 	& $512^2\times 128$ & 100 & $20$ &1.22 & $0.35$ &  0.040 & 0.0045 & 0.0032 & 0.007 & 0.0015\\    
\hline    
	SG-hydro 	& $512^2\times 128$ & 200 & 20 & 1.26 & 0.412 & 0.109 & 0 & 0.0059 & 0.0096 & 0  \\    
	SGMRI-20 	& $512^2\times 128$ & 320 & $20$ & $1.20$ &  0.362 & 0.041 & 0.0061 & $0.0034$ & 0.007 & $0.0017$  \\    
	SGMRI-10 	& $512^2\times 128$ & 100 & $10$ & $1.21$ &  0.363 & 0.062 & 0.021 & 0.0064 & 0.010 & 0.0071\\    
	SGMRI-5 	& $512^2\times 128$ & 60 & $5$ & $1.17$ &  0.33 & 0.067 & 0.043 & 0.011 & 0.015 & 0.014\\    
	SGMRI-2 	& $512^2\times 128$ & 14 & $2$ & frag &  frag & frag & frag & frag & frag & frag\\  
	\hline    
	SGMRI-20-Rm100 	& $128^2\times 64$ &  600 & 20 & 1.25 & 0.382 & 0.0259 & 0.272  & 0.0048 & 0.0025 & 0.021  \\   
	SGMRI-100-Rm100 	& $128^2\times 64$ &  400 & 20 & 1.25 & 0.49 & 0.014 & 0.14  & 0.0022 & 0.0017 & 0.0064  \\     
	SGMRI-20-By0.1 	& $512^2\times 128$ & 200 & $20$ & 1.29 &  0.40 & 0.042 & 0.014 & 0.0042 & 0.0065 & 0.011\\   
	SGMRI-20-Bz0.1	& $512^2\times 128$ & 40 & $20$ & $1.54$ &  0.58 & 0.113 & 0.368 & 0.036 & 0.0024 & 0.167\\ 
	\hline
\end{tabular}  
\vspace{0.5cm}
\caption{Simulations runs and their box and time-average turbulent quantities. The first three simulations (small box MRI without GI) have $L_x=2\, H_0, L_y=4\, H_0,$ while the rest have $L_x=L_y=20\, H_0$. The third column indicates the time over which quantities have been averaged (excluding transient phases). $Q$ is the Toomre parameter, $E_T$, $E_c$ and $E_m$ are respectively the internal, kinetic and magnetic energy, $H_{xy}$, $G_{xy}$ and $M_{xy}$ are the Reynolds, gravitational and Maxwell stresses. Note that for SGMRI-2, ``frag" means fragmentation.}  
\vspace{0.5cm}
\label{table1}
\end{table*}  

\begin{table*}    
\centering 
\begin{tabular}{c c c c c c c }          
\hline                        
Run & Time & $\tau_c$ & $\tau_w$ & $\dot{m}_w$ & $\alpha$  & $\alpha_{\text{th}}$\\   
 & ($\Omega^{-1}$) & ($\Omega^{-1}$) & ($\Omega^{-1}$)& & &  $=\tau_c^{-1}+\tau_w^{-1}$ \\   
\hline  
	MRI-S1 	& 1500 & $\infty$ & 148.7 & $2.9\times10^{-3}$ & 0.0076 & 0.0067\\   
	MRI-S2 	& 1500 & $\infty$ & x & x & 0.0075 & x\\  
	MRI-S3 	& 1200 & $\infty$ & x & x & 0.0068 & x\\  
	MRI-L1 	& 1500 & $\infty$ & 248.5 & $1.76\times10^{-3}$ & 0.0043 & 0.0040\\    
\hline
	MRISG-100 	 & 600 & $100$ & 233.7 & $7.09\times10^{-4}$ & 0.0133 & 0.0142\\    
\hline    
	SG-ref 	&  200 & 20 & 150.0 & $5.0\times10^{-4}$ & 0.056 & 0.057 \\    
	SGMRI-20 	&320 & $20$ & 110.6 & $7.46\times10^{-4}$ & 0.051 & 0.059 \\    
	SGMRI-10 	&  0 & $10$ & 85.4 & $9.3\times10^{-4}$ & $0.10$ &  0.11\\   
\hline
\end{tabular}  
\vspace{0.5cm}
\caption{Wind and transport quantities of some simulations shown in Table \ref{table1}. $\tau_w$ is the timescale at which total energy is lost through winds (defined in Eq.~\ref{eq_tw}), $m_w$ is the time-averaged mass loss rate through the vertical boundaries, $\alpha$ is the transport efficiency and  $\alpha_{\text{th}}$ is the theoretical efficiency given by the total averaged energy equation (\ref{energy_eq}).   Note $\alpha$ is defined as the standard ratio of stress over pressure without the factor $q\gamma$ and thus differs from \citet{Gammie2001} and \citet{riols17b}.}
\vspace{0.5cm}
\label{table2}
\end{table*}  
 \begin{figure*}
\centering
\includegraphics[width=0.98\textwidth]{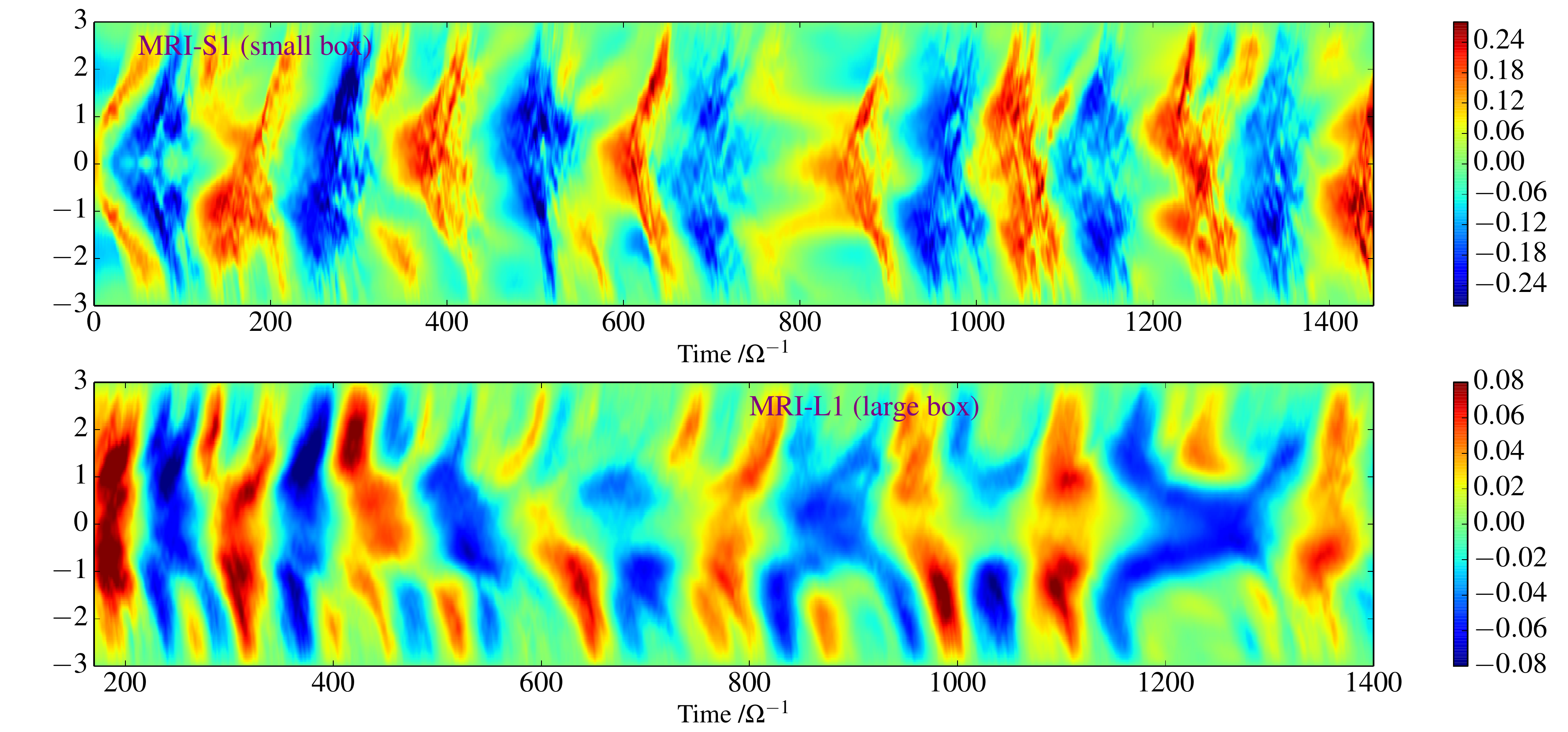}
\includegraphics[width=0.98\textwidth]{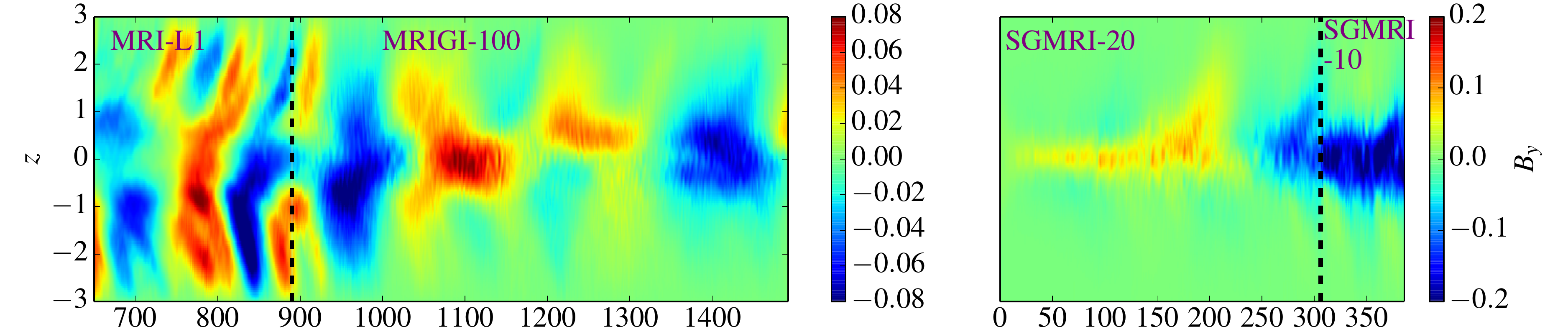}
 \caption{Space-time diagrams of the horizontally averaged $B_y$. Top
   panel: the small-box pure MRI simulation MRI-S1. Centre panel: the
   large-box pure MRI simulation MRI-L1. Bottom left: simulation
   MRI-L1 and the MHD gravito-turbulent simulation with
   $\tau_c=100\,\Omega^{-1}$ started from it (MRISG-100). The vertical
   dashed line indicates the transition from one to the other (at $t=t_2 \simeq 900\,\Omega^{-1}$) . Bottom
   right: MRI and self-gravitating simulations with
   $\tau_c=20\Omega^{-1}$ and $\tau_c=10\Omega^{-1}$, with the dashed
   vertical line again indicating the transition from the former to the
   latter.}
\label{fig_byzt}
 \end{figure*}
\begin{figure*}
\includegraphics[width=0.9\textwidth]{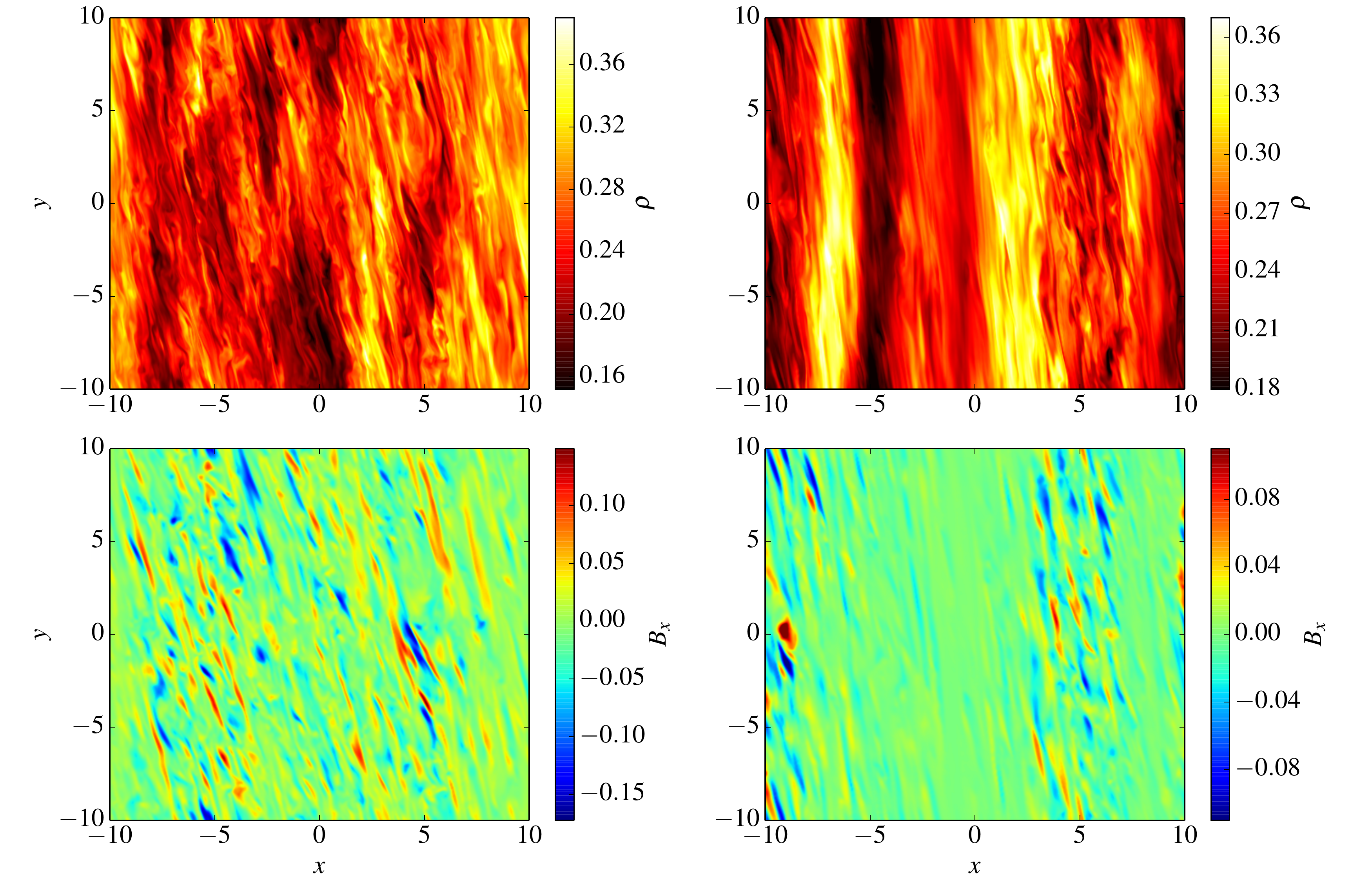}\vspace{0.8cm}
\caption{Snapshots of $\rho$ (top) and $B_x$ (bottom) in the large-box
  pure MRI
  simulation MRI-L1 at $z=H_0$ at
  two different times. The left panels are for $t=300\,\Omega^{-1}$ while the right panels are for $t=700\,\Omega^{-1}$ (when zonal structures are most pronounced).}
\label{fig_mri_rho}
\end{figure*}
\begin{figure}
\centering
\includegraphics[width=\columnwidth]{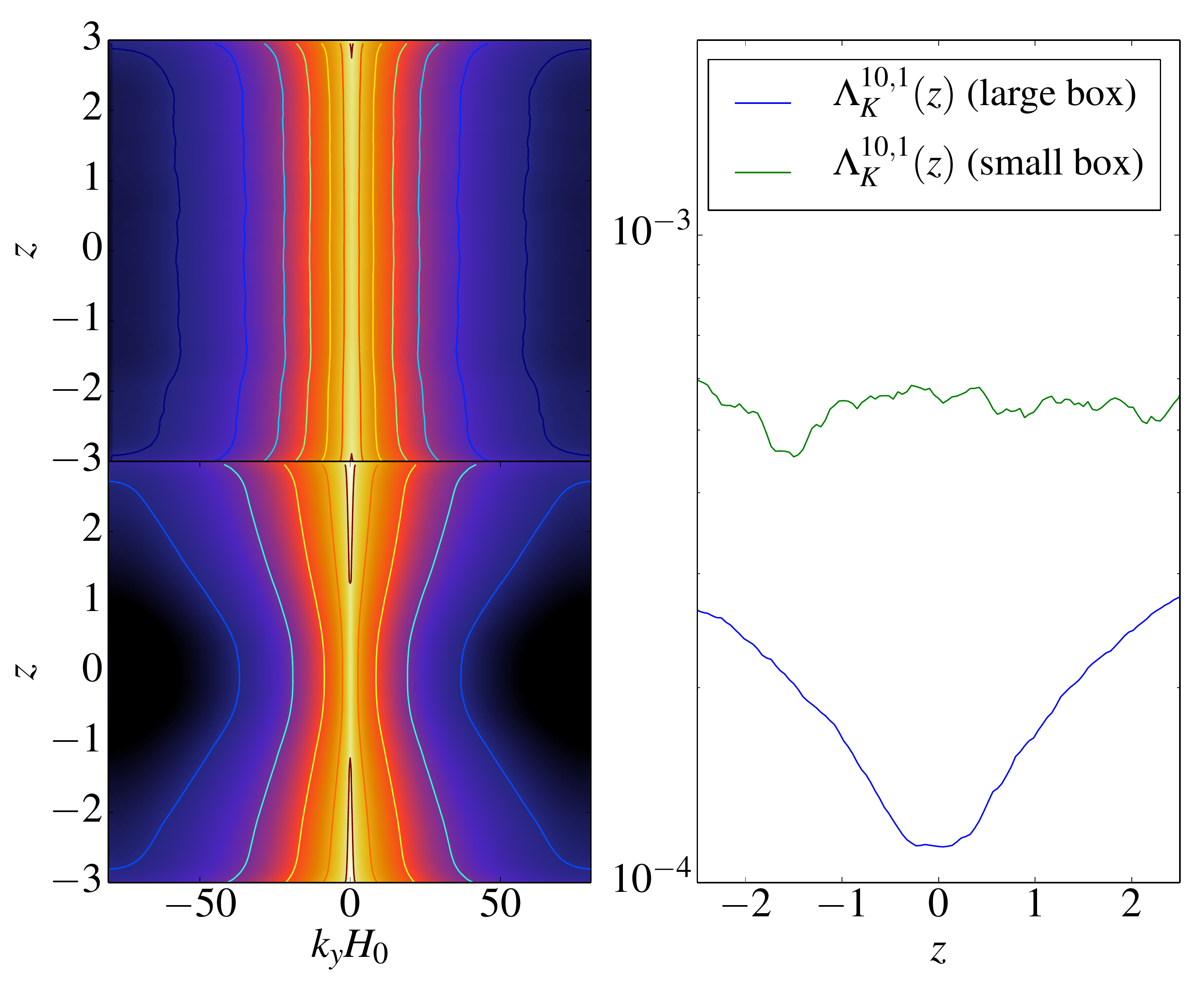}
 \caption{Left: colourmaps showing the time-averaged 1D kinetic power spectrum $\mathcal{E}_K(k_y,z)$ as a function of altitude $z$, for the small box (top) and large box (bottom) runs. On right, small to large scale ratios $\Lambda^{10,1}_{K}(z)$ between the PSD at $k_y=10\pi H_0^{-1}$ and $k_y=\pi H_0^{-1}$ (green curve for small box and blue curve for large box).  }
\label{fig_mri_spectrum}
 \end{figure} 
\begin{figure}
\centering
\includegraphics[width=\columnwidth]{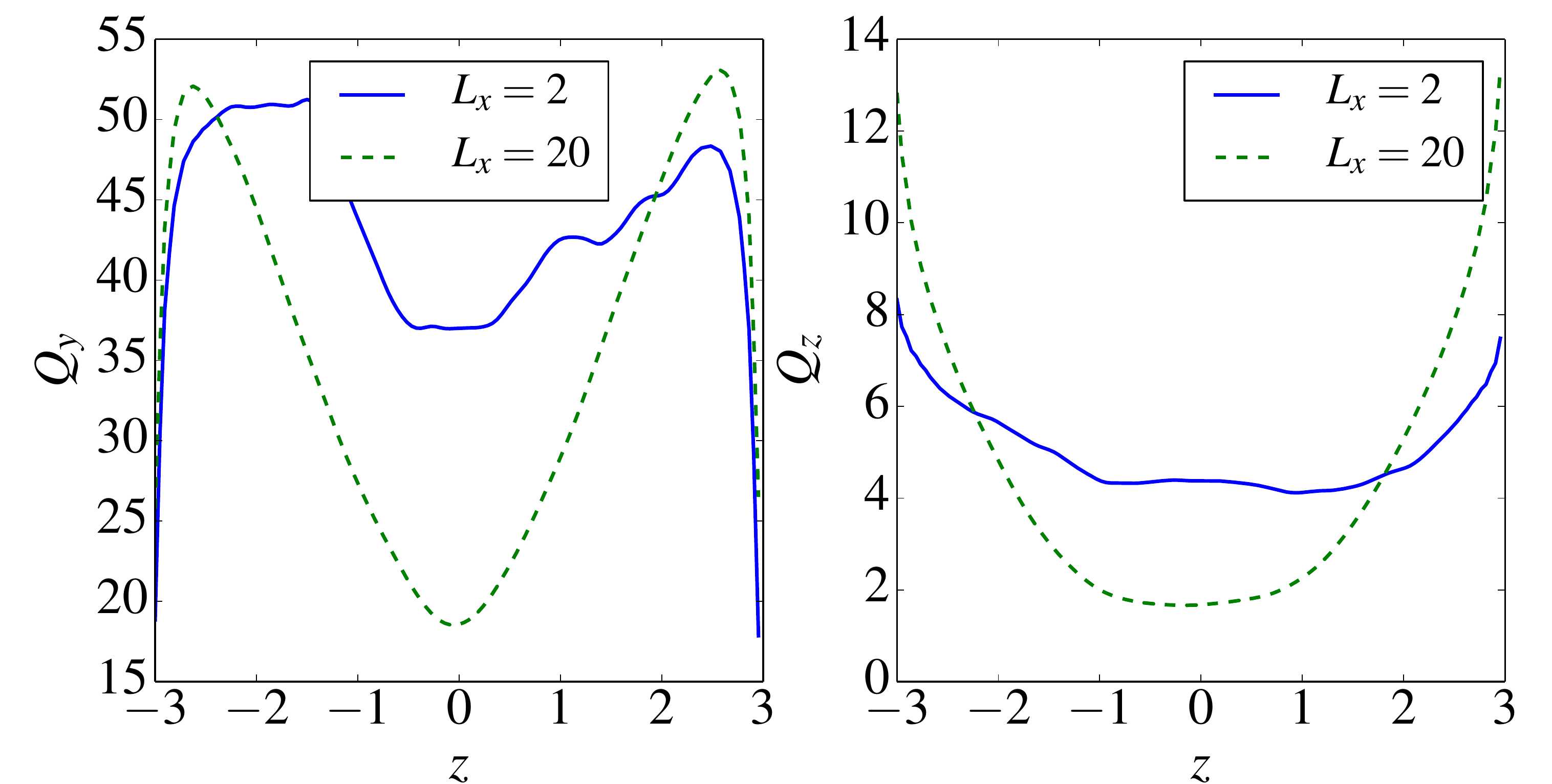}
 \caption{Quality factors in the $y$ (left) and $z$ direction (right). Blue/plain curves are computed from the small box MRI simulation (MRI-S1) while green/dashed curves are from the large box MRI simulation (MRI-L1). Both simulations have a resolution of $26$ points per disc scale height in horizontal directions ($\Delta x = 0.039$) and $22$ points per $H_0$ in the vertical direction ($\Delta z = 0.046$)}
\label{fig_qfactor}
 \end{figure} 

\subsection{Small boxes}

We performed a first test simulation of stratified MRI-driven
turbulence, labelled MRI-S1 (see Table \ref{table1}) with $\gamma=5/3$
and no cooling ($\tau_c= \infty$).  This simulation has zero-net flux,
no explicit diffusion, no self-gravity, and has been run in a small
box of size $L_x=2\,H_0$, $L_y=4\,H_0$, $L_z=6\,H_0$ for $1500
\,\Omega^{-1}$.  The resolution is 26 points per $H_0$ in $x$ and $y$,
and 21.3 points per $H_0$ in the vertical direction (our standard
resolution for most of the simulations in this paper). Actually this  
setup is similar to the stratified simulations of \citet{simon11}
except that $\gamma$ differs
 from 1.

Figure \ref{fig_mri_small} shows the time evolution of $\alpha$, the
ratio of total stress over pressure. The turbulent activity is
sustained during the whole simulation and saturates at
$\alpha=0.0076$. This value is similar to that found by
\citet{gressel13} but slightly smaller than those found in isothermal
simulations of similar resolution per $H_0$:
\citet{davis10} found $\alpha \simeq 0.01$ while \citet{simon11} found
$\alpha\simeq 0.027$ with no explicit diffusion. The ratio of Maxwell stress to
Reynolds stress $M_{xy}/H_{xy}\simeq  3.32$ is consistent with
previous stratified MRI simulations.  

The space-time diagram shown in Fig.~\ref{fig_byzt} (top panel) reveals that
the large-scale toroidal field oscillates between positive and
negative values with a period of $\sim 20-25$ orbits. Each reversal
starts in the midplane and propagates upwards/downwards into the disk 
atmosphere, thus generating ``butterfly diagrams".  This behaviour has
appears in nearly all stratified MRI simulations
\citep{branden95,davis10,simon11,oishi11,gressel15} and is classically
attributed to a large-scale dynamo driven by non-axisymmetric
MRI waves \citep{rincon07b,lesur08}. Note that the period of each
reversal in our simulations is twice longer than that typically inferred
from classical isothermal simulations. Moreover, the butterfly
diagram is slightly erratic: e.g.\ the toroidal field in the
midplane sometimes conserves its polarity (as between $t=300$ and
$t=450\,\Omega^{-1}$). \citet{bodo12} and \citet{gressel13} also
reported a marked and analogous sensitivity of the butterfly patterns to the
thermodynamics (and vertical boundary conditions), while \citet{hirose14} and \citet{coleman17} suggest that hydrodynamic mixing of magnetic fields by convective motions might
weaken the field reversals. We, however, find no evidence
of convection in our simulations, although the squared Brunt-Vaisala 
frequency $N^2$ occasionally takes small negative values near the
midplane. 

\subsection{Large boxes and zonal flows}
\label{large_box}

Since our aim is to capture both MRI and GI, simulations have to be
run in much larger boxes than used in MRI-S1. Therefore we ran a
second test MRI simulation without self-gravity, labelled MRI-L1,
in a box 10 and 5 times larger in  $x$ and $y$ respectively, 
but with the same resolution in both directions (26 points per $H_0$).\\

Table \ref{table1} contains lists of time-averaged quantities
that may be compared with those of MRI-S1. We found that the Reynolds and
Maxwell stresses are both smaller in the larger box as compared with
the small box. The magnetic
energy halves, although kinetic and internal energy
remain similar.
 
The space-time diagram of Fig.~\ref{fig_byzt} (centre
panel) shows that the large-scale dynamo field $B_y$ reverses fairly
regularly every $\sim 5-10$  orbits. However, the turbulent flow undergoes long term
variability ($\simeq 50$ orbits) with periods
of high/moderate magnetic activity (e.g.\ in the early stage of the
simulation between $t=200$ and $450 \,\Omega^{-1}$ and also around $t=1000
\,\Omega^{-1}$) followed by periods of weaker activity
(between $t=500$ and $800\, \Omega^{-1}$). The polarity of the magnetic
field during the latter phase can be markedly asymmetric about the
midplane. 

To better understand the origin of these variations, we
plot in Fig.~\ref{fig_mri_rho} the density $\rho$ and the radial
field $B_x$ in a plane $z=H_0$, at two different epochs,
$t=300\,\Omega^{-1}$ (corresponding to high activity)  and
$t=700\,\Omega^{-1}$ (corresponding to low activity).  In the first
case (left panels), the turbulence is well-developed, homogeneous, and
mainly small-scale.  Magnetic bundles of size $\sim 0.2 \,H_0$ are
elongated along the shear and distributed uniformly. In the
second case (right panels), the turbulence is weaker
and more
patchy, and the flow has formed into larger scale structures.
In particular, the density
develops long-lived axisymmetric rings or ``zonal flows". The density
contrast between each band is significant, of the order $50\%$ of
the background. Small-scale magnetic filaments are still
visible but they remain confined to azimuthal bands,
where the Alfv\'en speed is larger. Note that at $t=1000
\,\Omega^{-1}$, when the magnetic activity regains its strength,  
the density is dominated again by small-scale structures, 
although faint zonal flows can still be distinguished in the background. \\

MRI zonal flows have been reported in several MRI
simulations with radial box sizes much larger than $H_0$
\citep{johansen09,simon12,kunz13,bai14b}. Their provenance remains
unclear and
a subject of ungoing research.
Usually the lengthscale of these
features increases to a value near the radial box size, and it is likely
that finite-domain effects rather than physical effects limit this growth.
In keeping with previous results our large box ($L_x=L_y=20\,H_0$)
hosts at least two zonal
structures.

 The emergence of large-scale zonal flows can
have a considerable impact on the overall dynamics. First, a significant amount of
angular momentum may be transported on scales $\gg H_0$
\citep{beckwith11,simon12}, which breaks the assumption of locality
and casts doubt on the validity of the alpha model. However, we found
the opposite:
large-scale zonal flows induce a drop in the total
box-averaged stress and transport. Second, zonal flows strongly influence the
turbulent spectrum and small-scale structure.
Figure \ref{fig_mri_spectrum} shows the 1D kinetic power spectrum
$\mathcal{E}_K(k_y,z)$, averaged over $k_x$, as a function of $z$ and
$k_y$. 
In the small box, kinetic energy is spread uniformly in $z$ over
$k_y$. In contrast, the spectrum in the large-box simulation
adopts 
a funnel
shape. More energy is found on large scales and power is depleted on
small scales in the midplane region, where the zonal flows are
present. This behaviour is more obvious in plots of the
small/large-scale ratio $\Lambda^{10,1}_{K}(z)$ (right panel). 
An increased box size makes MRI-driven flows more
laminar in the midplane. Indeed, $B_y$ and the Maxwell stress
$-B_xB_y$ are smaller in the midplane than in the
corona. One way to explain this result is that zonal flows pump energy
from small scales to large scales \citep{simon12}. A more worrying
possibility could be that these zonal flows produce regions of very
low magnetization, in which the MRI grows only on scales  
shorter than the grid size and is hence misrepresented in the
simulation. 
This prospect is discussed in further detail in the
next subsection. 

In conclusion, the emergence of zonal flows in large boxes, residing 
preferentially in the midplane regions, seems to induce a 
drop in turbulent activity and the formation of ``deserts" in which
small-scale turbulence is absent.

\subsection{The resolution problem}
\label{convergence_MRI}

If the dependence of the MRI on box size is
an issue, then what of its dependence on resolution?  Is MRI
turbulence adequately resolved with 26
points per $H_0$ in $x$ and $y$ and 21 points in $z$? The question of
the convergence of MRI with resolution in stratified shearing box
simulations has been debated for a number of years. Early simulations by
\citet{shi10,davis10,simon11} suggested convergence of $\alpha$ with
resolution. However, the most recent and numerically intensive simulations by
\citet{bodo14} and \citet{gammie17}, 
indicate the contrary: convergence is still not 
obtained even up to 256 points per $H_0$ (with $\alpha \sim N^{-\frac{1}{3}}$). \\

As no convergence test has been performed for MRI in the diabatic
case, we examined the properties of the saturated state at
larger resolution. For computational reasons, we restricted our study
to the case of a small box. Starting from the same initial condition
as MRI-S1, we performed  two simulations, one with resolution of 32
points per $H_0$ in $x$ and $y$ but the same resolution in $z$
(MRI-S2), the second with double resolution in each direction (MRI-S3,
64 points per $H_0$ in $x$ and $y$ and 43 points per $H_0$ in
$z$). Each simulation was run for more than 1000 $\Omega^{-1}$ to
obtain meaningful statistical averages. Tables \ref{table1} and
\ref{table2} show that all turbulent average properties, and  in
particular $\alpha$, slightly decrease when resolution is increased
although no real trend can be inferred from our runs. According to
\citet{gammie17}, a significant change in average quantities would
require us to go beyond 64 points per $H_0$ (which means at least
$2000\times 2000\times 512$ in the large boxes we intend to use). 
The main point we wish to make is that numerical convergence with
resolution cannot be achieved with our resources and is unlikely to
exist in any case.

Though the MRI in our simulations may be formally unresolved, its
interaction with GI may still be adequately described.
To explore this we next consider simulations with standard resolution
(26 points per $H_0$) and compare the characteristic MRI wavelength
(for which the MRI growth rate is maximum) with the grid size. The
ratio between these two lengthscales,  
 called the quality factor, is approximately \citep{sano04}:
\begin{equation}
Q_i(z)= \dfrac{2\pi \overline{v_{A}}_i(z)}{\Omega \Delta x_i}
\end{equation} 
where $v_{{A}_i}=B_i/\sqrt{\rho}$ denotes the Alfv\'en speed in the
direction $i=(x,y,z)$. This number is obviously rather crude but gives
a feeling for how well the largest MRI modes are resolved. 

 Figure \ref{fig_qfactor} shows the vertical profiles of $Q_y$ and $Q_z$,
averaged in time over 200 orbits, for the small and large box
simulations. In the midplane, $Q_y \gtrsim  20$ for both simulations,
taking larger value in the upper layers, which means that the
non-axisymmetric MRI modes, supported by the toroidal field, might be
adequately resolved in both cases. However, we note that $Q_y$ drops by
a factor 2 compared to the small box run, in particular in the
midplane. This is due to the zonal flows discussed in
Section \ref{large_box}. In fact, the situation is worse than 
suggested by Fig.~\ref{fig_qfactor} because the quality factors shown are
horizontally averaged. In fact, the zonal flows produce
weakly
magnetised bands in $x$ in which locally $Q_y \simeq 1-2$, very low
values indeed. 
In these weakly magnetised regions, one would need to double or
quadruple the resolution to resolve the most unstable MRI modes,
although it is not guaranteed even then that these structures would
maintain the same level of magnetization when the resolution is increased;
the magnetisation might fall with increasing resolution
\citep{gammie17}.

MRI modes supported by the vertical field are even worse and are only marginally
resolved on the average, especially in the midplane where $Q_z\simeq 2 $ or
$Q_z\simeq 5 $  respectively for $L_x=20$ and $L_x=2$. That said, the issue of
vertical resolution is probably of less importance as $B_z$ fluctuates
rapidly and is mainly small scale,  
and thus is unlikely to support a coherent MRI mode \citep{simon11}. \\

In conclusion, our standard resolution (26 and 22
points per $H_0$ in the horizontal and vertical direction
respectively) is unconverged, a problem that we must make
explicit at this point. Although a resolution of
64 points per $H_0$ does not
seem to drastically change  the average saturated state, the very
small-scale MRI ($\lll H_0$) is probably misrepresented.  In
particular, the generation of large-scale zonal flows, combined with a
lack of resolution, weakens MRI activity in the midplane.
 As suggested by \citet{gammie17}, the worst case scenario is that
magnetisation slowly decreases forever with resolution. This would
cast serious doubt on the MRI viability in zero-net-flux configuration
without explicit diffusion. While we acknowledge these problems, we do
 believe that out setup is
probably sufficient to capture the most unstable modes
on intermediate scale, as well as the main nonlinear properties of the
MRI dynamo. Most importantly, we can still learn a great deal from the
competition of gravitoturbulence and the zero-net-flux MRI, 
even if the latter suffers from the problems
described above.

\section{MHD simulations with self-gravity}
\label{sec_GIMRI}

We are now in a position to analyse the interaction between GI and the
MRI, and more generally between 3D gravitoturbulence and zero-net-flux
magnetic fields. Our first set of runs examines long cooling
times $\geq 100\, \Omega^{-1}$  for which GI and non-axisymmetric MRI
are expected to be of similar intensity (see Section
\ref{choice_cooling}).  The second set corresponds to an intermediate
cooling time $\tau_c= 20\,  \Omega^{-1}$. We also compare
states initialized from pure MRI turbulence to those initialized
from pure hydrodynamic gravitoturbulence, so as to rule out any
dependence on the initial condition.
Our third set of simulations explores the low cooling time
regime where GI is especially strong and fragmentation can occur. 
Lastly, we present simulations with an imposed
magnetic field. 

\subsection{A matter of cooling times}
\label{choice_cooling}

Unlike the MRI runs of Section \ref{sec_MRI}, we introduce a
cooling law that favours ``gravito-MRI" states.  The cooling time
$\tau_c$ turns out to be the key control parameter here: small
$\tau_c\simeq \Omega^{-1}$ produces vigorous GI turbulence,
or fragmentation in the extreme case. In
the opposite limit, inefficient cooling $\tau_c \gg 100 \, \Omega^{-1}$
weakens gravito-turbulent activity and thus sets the scene for MRI
to dominate. As a first step, we study the case for which MRI and GI have comparable strength. 

We consider the two instabilities separately and
arrange for a situation where the angular momentum transport
associated with each are roughly equal:
\begin{equation}
\alpha_{\text{MRI}}\simeq\alpha_{\text{GI}}.
\end{equation}
The MRI transport efficiency is given in Table \ref{table1}
($\alpha_{\text{MRI}}\simeq 0.0043$) while GI efficiency is known to
be inversely proportional to the cooling time, following
\citet{Gammie2001}. This gives a first estimate for $\tau_c$
 that allows both instabilities to be of similar magnitude 
\begin{equation}
\label{eq_tau1}
\tau_c \simeq \frac{1}{q\Omega (\gamma-1)\alpha_{\text{MRI}} } \simeq 230\, \Omega^{-1}.
\end{equation}

This relation, however, is inaccurate when the two
instabilities are both operating, and excludes wind cooling. 
Let us assume a hypothetical
ideal case where each instability does not affect the other, 
and thus both contribute to the transport
and disc heating additively. The relation given by Eq.~\eqref{energy_eq}, based on
energy conservation, applies now to the full system \{GI+MRI\} which
has two sources of heat. If we denote by $\tau_w$ the timescale of energy
loss through winds, the energy balance is given by
\begin{equation}
\frac{1}{\tau_c}+\frac{1}{\tau_w}=q(\gamma-1)(\alpha_{\text{GI}}+\alpha_{\text{MRI}}).
\end{equation}
One needs to estimate $\tau_w$. One possibility is that
$\alpha_{\text{MRI}}=1/\tau_w$ which means that the disc winds are not
affected by GI, and therefore the estimate given by Eq.~\eqref{eq_tau1}
remains valid. The other possibility is that the wind becomes negligible when GI and MRI coexist; in that case we find that 
\begin{equation}
\label{eq_tau2}
\tau_c \simeq \frac{1}{2q\Omega (\gamma-1)\alpha_{\text{MRI}} } \simeq 115\, \Omega^{-1}
\end{equation}
In summary the critical cooling time at which the MRI and GI are
equally strong is $\sim 100\Omega^{-1}$. For cooling times less than
this critical value, GI should dominate. 

\subsection{Regime of inefficient cooling ($\tau_c \geq 100\,\Omega^{-1}$)} 

We begin by examining the regime of long cooling times in which the
MRI and GI are roughly comparable. We treat two cases
$\tau_c=100\Omega^{-1}$ and $\tau_c=200\Omega^{-1}$. 

\label{sec_MRIstart}
\begin{figure*}
\centering
\includegraphics[width=0.8\textwidth]{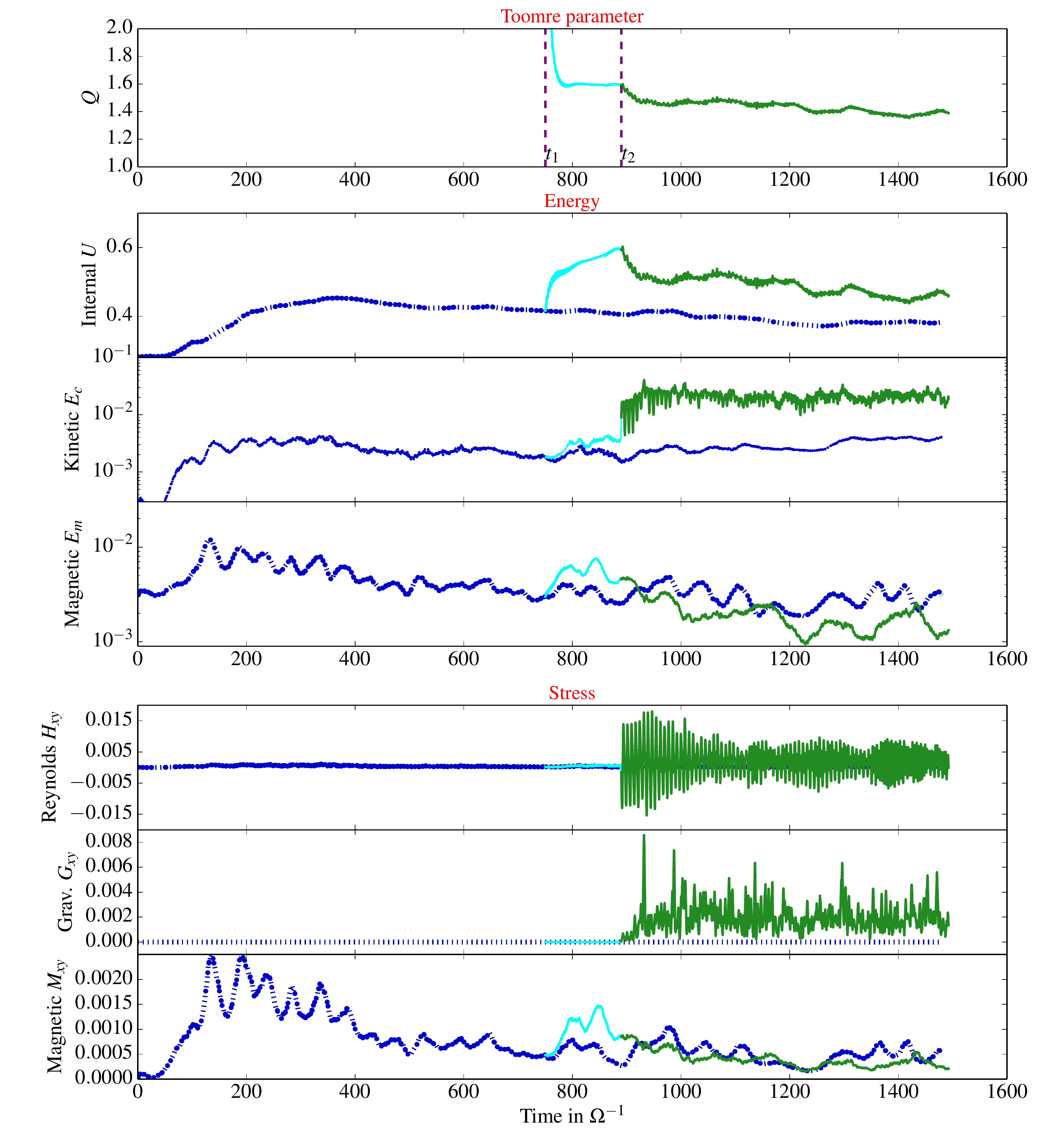}
 \caption{Time-evolution of various quantities, averaged over a box of
   size $L_x=20$, $L_y=20$ and $L_z=6 \,H_0$. From top to bottom,
   Toomre parameter $Q$, box average internal, kinetic and magnetic
   energy, box average   Reynolds, gravitational and Mawxell
   stress. The  blue/dashed curves corresponds to the MRI run without
   self-gravity (MRI-L1) while the green/plain curves represent the
   run with self-gravity and $\tau_c=100\, \Omega^{-1}$
   (MRISG-100). The cyan/light curve represents the transition phase
    in which only the mean vertical component of self-gravity is
   incorporated (no GI fluctuations). All simulations have a resolution of $512\times512\times 128$.}  
\label{fig_average}
 \end{figure*}
 \begin{figure*}
\includegraphics[width=\textwidth,trim=0cm 1.cm 0cm 0.5cm, clip=true]{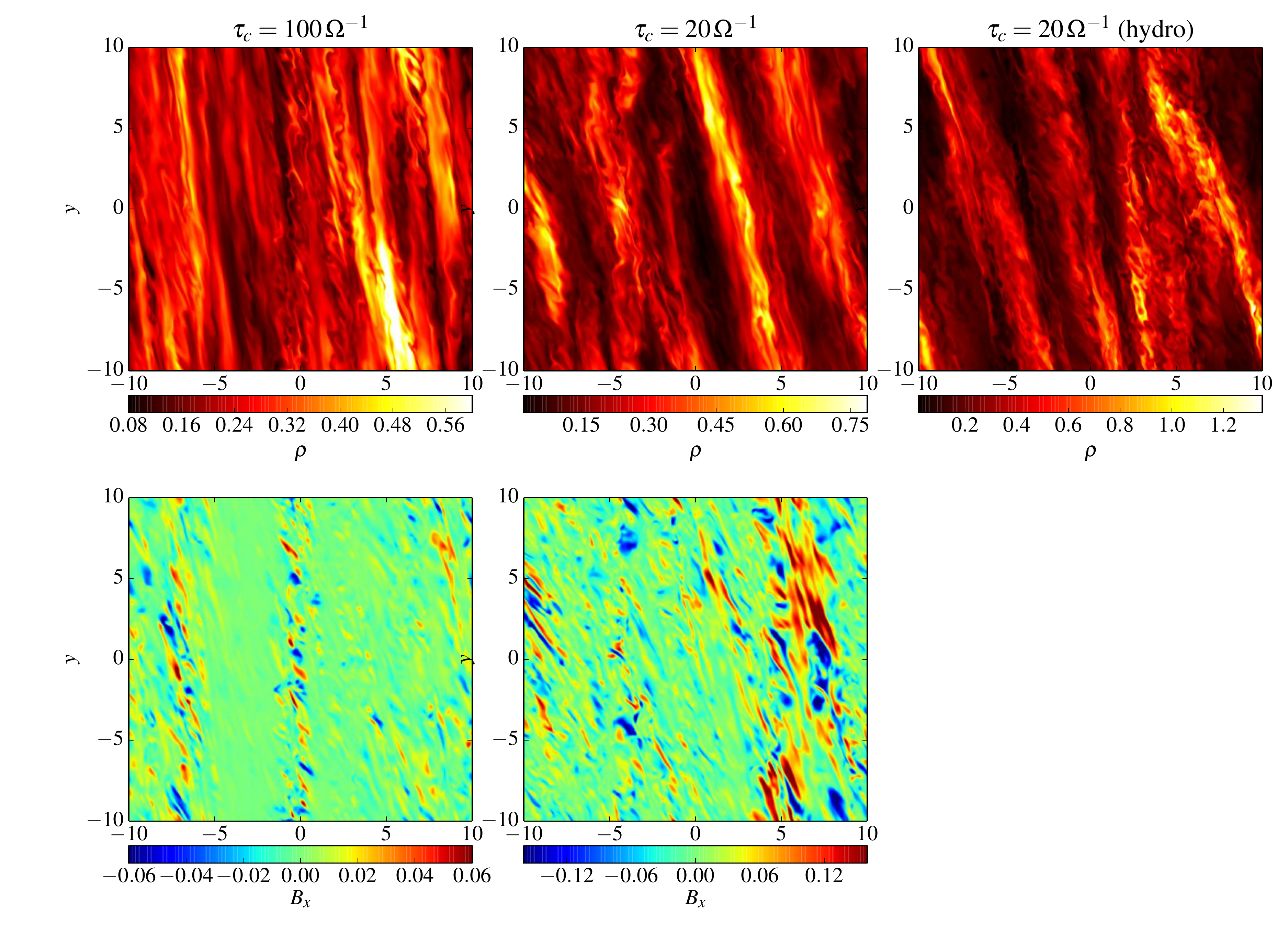}
\includegraphics[width=0.99\textwidth,trim=0cm 0.cm 3cm 2cm, clip=true]{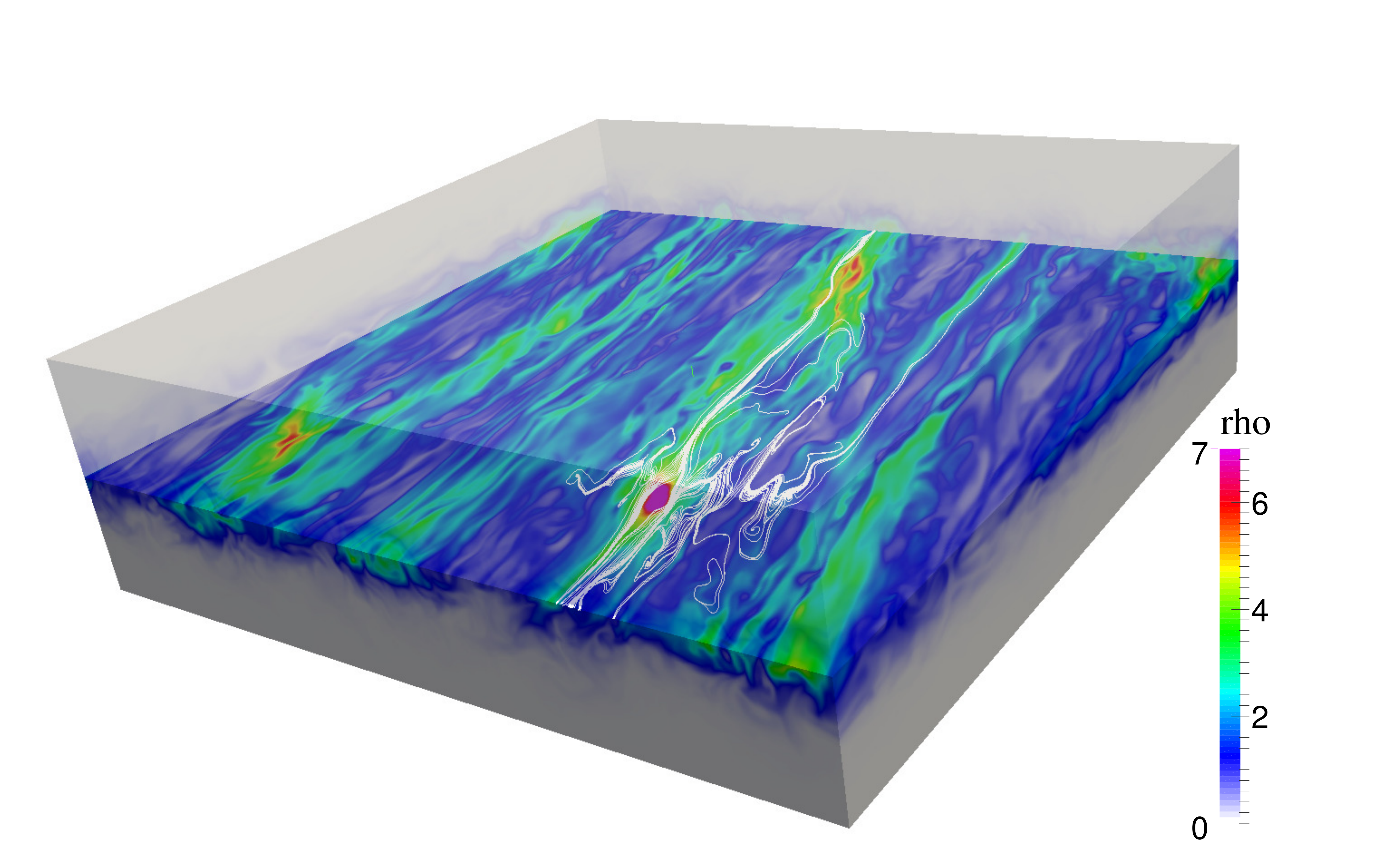}
\caption{Top: snapshots of $\rho$ and $B_x$ at $z=H_0$, taken from
  different simulations. From left to right, magnetized GI turbulent
  states with $\tau_c=100\,\Omega^{-1}$ (MRISG-100),
  $\tau_c=20\,\Omega^{-1}$ (MRISG-20) and a pure hydrodynamic GI state
  (SG-hydro) with $\tau_c=20\,\Omega^{-1}$. Bottom: 3D view of a
  plasmoid embedded in a magnetic island from a simulation with $\tau_c=5\Omega^{-1}$. The colours indicate the density and white lines represent some horizontal magnetic field lines. All simulation have  $L_x=L_y=20\,H_0$ and a resolution of $512\times512\times 128$. }
\label{fig_SGmri_rho}
\end{figure*}
\begin{figure*}
\centering
\includegraphics[width=1\textwidth]{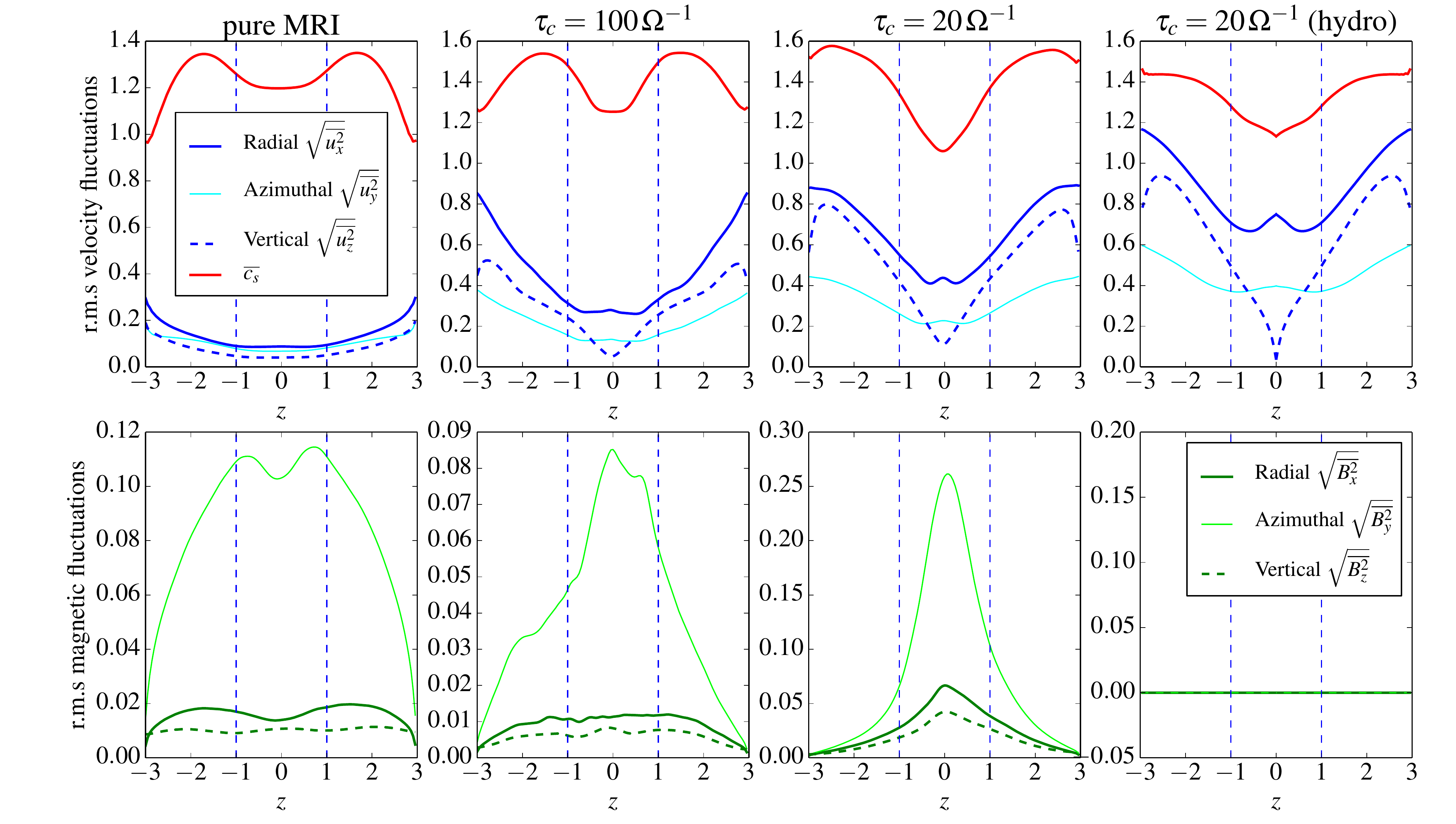}
 \caption{Vertical profiles of turbulent r.m.s velocity (top) and magnetic (bottom) components, time averaged over four different simulations. From left to right, the pure MRI state without self-gravity (MRI-L1), the combined MRI$+$GI state with $\tau_c=100\,\Omega^{-1}$ (MRISG-100), the case of intermediate $\tau_c=20\,\Omega^{-1}$ (SGMRI-20) and the pure hydrodynamic state (SG-hydro) with $\tau_c=20\,\Omega^{-1}$.}
\label{fig_vfluct}
 \end{figure*}
 \subsubsection{Initial state and simulation timeline}
\label{init_timeline}

Our first runs start from a fully developed MRI turbulent state with
$Q\rightarrow \infty$ taken from the large-box simulation MRI-L1 at
$t_1=750\,\Omega^{-1}$. Initially we keep $\tau_c=\infty$ and
do not introduce full self-gravity straight away, but only
its mean and static vertical component. The reason is to check
that MRI can be sustained in a disc compressed by its own gravity,
neglecting the action of GI fluctuations and spiral waves. To avoid
sharp changes to the disc structure and thermodynamics, the Toomre
parameter $Q$ is progressively decreased and set to a value around 1.6
(see top panel of Fig.~\ref{fig_average} from $t_1=750\,
\Omega^{-1}$). This is done by taking $G \propto
1-e^{-(t-t_1)/\tau_G}$ with $\tau_G\simeq 50 \,\Omega^{-1}$. \\

By
$t\simeq 800\, \Omega^{-1}$, the disc has converged upon a new turbulent
state whose averaged properties are plotted in light/ cyan in
 Fig.~\ref{fig_average}. Note that only 20 orbits (between $750$ and
$900\,\Omega^{-1}$) are represented here to avoid plot overloading,
but we actually obtained this state for longer time. Finally at
$t=900\, \Omega^{-1}$, full self-gravity (including its fluctuating
part) and cooling are introduced.  The evolution of
corresponding averaged quantities are plotted in green in
Fig.~\ref{fig_average} for the case $\tau_c=100 \,\Omega^{-1}$. They
can  be directly compared to those obtained in the pure MRI simulation
(blue curves, MRI-L1). 
 The new state with self-gravity is labelled ``MRISG-100".

\subsubsection{MRI and its interaction with GI}
\label{MRIGIsim}

Fig.~\ref{fig_average} shows that during the transition phase (between
$t_1$ and $t_2$, cyan/light curves), when only the mean vertical
component of self-gravity is considered, the internal energy slightly
increases from 0.4 to 0.6 due to the disc compression, but the other
quantities do not change all that much. The turbulent stresses, 
normalized to the pressure, are comparable to those
obtained in the limit $Q\rightarrow \infty$. The magnetic field
reverses 3-4 times and we checked that 
 the butterfly diagram is not affected; a fraction of the toroidal
 magnetic flux is transported upward by buoyancy, while the 
dynamo cycle period is similar to the case $Q\rightarrow \infty$.\\

At $t_2=900\,\Omega^{-1}$, as full self-gravity is introduced, the
turbulent state changes radically. Figure \ref{fig_average}
(green curves) shows that for $\tau_c=100 \,\Omega^{-1}$, the
mean Toomre parameter and internal energy drop under the effect of the
cooling but seem to both converge to a steady value (in particular $Q
\sim 1.4$) as soon as spiral shocks develop. Kinetic energy $E_c$ and
Reynolds stress $H_{xy}$ increase by a factor 10 while magnetic energy
and Maxwell stress slightly decrease but remain of the same order of
magnitude ($E_m$ decreases
actually by a factor $\simeq 2$ on average between $t=1000$ and
$t=1500\, \Omega^{-1}$). In addition to the Maxwell and Reynolds
stresses, the flow is subject to a  strong gravitational stress
$G_{xy} \simeq H_{xy}$. 
 Note that, unlike magnetic quantities, both hydrodynamical and gravitational turbulent components are highly fluctuating. \\

The important result here is that $E_m \ll E_c$ and $M_{xy} \ll G_{xy}
$.  
The gravitational stress, which is directly related to GI turbulence, is on
average 6 times larger than the Maxwell stress, which we associate at
this point with the MRI. The transport is then mainly driven by the
gravitational instability.  This might be a surprise because our cooling
time had been explicitly chosen in order to satisfy  $G_{xy} \simeq
M_{xy}$ (see Section \ref{choice_cooling}). Note that in making this
choice, we assumed that both instabilities do not interact with each
other, evidently an assumption that is incorrect. Moreover, as shown in
Table \ref{table2}, winds carry a non-negligible amount of energy (mainly internal energy).  
According to \ref{choice_cooling}, one may
argue that a cooling time of $100\,\Omega^{-1}$ is then still too low
to ensure $G_{xy}\simeq M_{xy}$. However, we checked that for a cooling time
$\tau_c=200 \,\Omega^{-1}$, the saturated state is in fact comparable
 (see Table \ref{table1}). It is unproductive to go to longer $\tau_c$:
 as we approach the limit $\tau_w \approx \tau_c$,
 the wind will cool the disc at a rate greater than the explicit Newtonian
 cooling.\\

Figure \ref{fig_SGmri_rho} (top left panels) shows a snapshot of the
density $\rho$ and radial magnetic field $B_x$ at $z=H_0$ for
$\tau_c=100\,\Omega^{-1}$. Surprisingly, the density field is not
dominated by large-scale spiral waves but rather by thin wispy
filaments, elongated in the radial direction and perturbed by
small-scale non-axisymmetric wobbles. We show in Section
\ref{transition_dynamo} that these small-scale features are probably
manifestations of a ``sluggish" MRI that persists in the
gravito-turbulent background. The manifestation of 
large-scale zonal flows in the density seem to have disappeared. 
But the bottom panel shows that the magnetic
field concentrates into small-scale bundles (of size $\ll H_0$)
localized preferentially along thin filaments or axisymmetric rings
(parallel to the $x$ axis). Although is it difficult to
determine the origin (MRI or not) of these structures, 
 they are reminiscent of those found in large box MRI. \\

Finally, we examined the evolution of the mean toroidal magnetic field
$B_y$. The space-time diagram of Fig.~\ref{fig_byzt} (bottom left)
shows that once self-gravity is included (at $t=t_2=900\,
\Omega^{-1}$), the field still reverses quasi-periodically, suggesting
the existence of a large-scale dynamo. However, the period of the
reversal is longer than in the pure MRI case (50 orbits instead
of 20) and the reversals are even more irregular. 
In addition, the butterfly patterns
disappear and the magnetic flux remains confined near the
midplane. The fact that magnetic flux cannot easily rise is possibly
due to the strong stratification in runs with self-gravity. Indeed, we
checked that the Brunt-Vaisala frequency increases rapidly with
$z$ in comparison to the case without GI. According to the Newcomb
criterion, magnetic buoyancy should be impeded. This behaviour,
however, may not be characteristic of 
realistic disc models with more sophisticated cooling treatments.

The confinement of the magnetic fluctuations can be also observed in
Fig.~\ref{fig_vfluct} (first and second bottom panels), which compares
the time-integrated r.m.s.\ magnetic components of MRI-L1 (pure MRI
without self-gravity) with that of MRISG-100 (with self-gravity). Although the
maximum amplitude of each component and the ratio $B_y/B_x$ are very
similar, their distribution along $z$ differs considerably. 
In the pure MRI case, $B_x$ and $B_y$ are distributed over a wide
range of altitude and are maximum at $z\simeq 1-2 H_0$. In the
self-gravitating case, they preferentially peak in the midplane. Note
also that velocity components are  
much stronger, in particular $v_x$ and $v_z$ in the corona (top panels of Fig.~\ref{fig_vfluct}).

\subsection{Regime of moderate cooling  ($\tau_c = 20 \,\Omega^{-1}$)}
\label{intermediate_cooling}
\subsubsection{Starting from an MRI-turbulent state with $Q\rightarrow \infty$}
\label{starting_MRI}

We performed a simulation, labelled MRISG-20, using the same
initialization as in Section \ref{init_timeline}, i.e.\ starting with an
MRI-turbulent state. The difference is that $\tau_c=20\, \Omega^{-1}$,
instead of 100 or 200. Table \ref{table1} shows that in this
intermediate regime, the activity is 2-3 times greater than for
$\tau_c=100\,\Omega^{-1}$ but the ratio between the Maxwell and
gravitational stresses and $E_m/E_c$ remains relatively small. There
is also a substantial drop in $Q$ and internal energy. Figure
\ref{fig_SGmri_rho} shows that large scale spiral waves,
characteristic of GI, become prominent, as opposed to the case
$\tau_c=100\,\Omega^{-1}$. The magnetic field forms small-scale
bundles with sizes comparable to those found at larger cooling
times. 
Magnetic structures either follow the spiral waves shape or regroup into radial axisymmetric bands. 

\subsubsection{Starting from hydrodynamic gravito-turbulence}

\begin{figure*}
\centering
\includegraphics[width=0.8\textwidth]{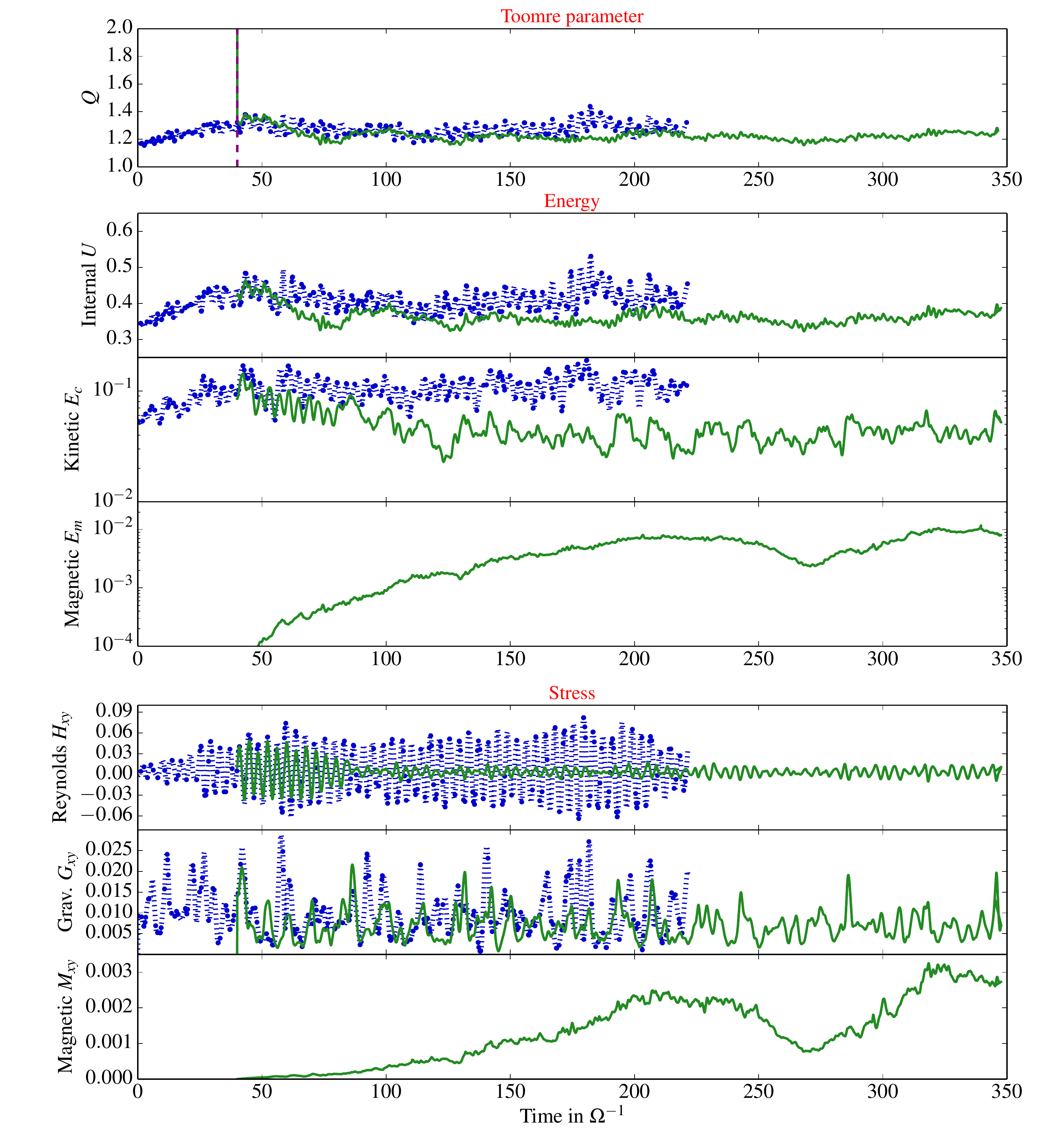}
 \caption{Time-evolution of various quantities, averaged over a box whose size is $L_x=20$, $L_y=20$ and $L_z=6 \,H_0$. From top to bottom, density-weighted average Toomre parameter $Q$, box average internal, kinetic and magnetic energy, box average   Reynolds, gravitational and Maxwell stress. The  blue/dashed curves corresponds to the pure 3D hydrodynamical gravito-turbulent state (SG-hydro) while the green/plain curve represents the same state with magnetic field (SGMRI-20), initialized from the hydrodynamic simulation. Simulations have a resolution of $512\times512\times 128$ and $\tau_c=20\, \Omega^{-1}$ .}
\label{fig_average2}
 \end{figure*}
 \begin{figure}
\centering
\includegraphics[width=\columnwidth]{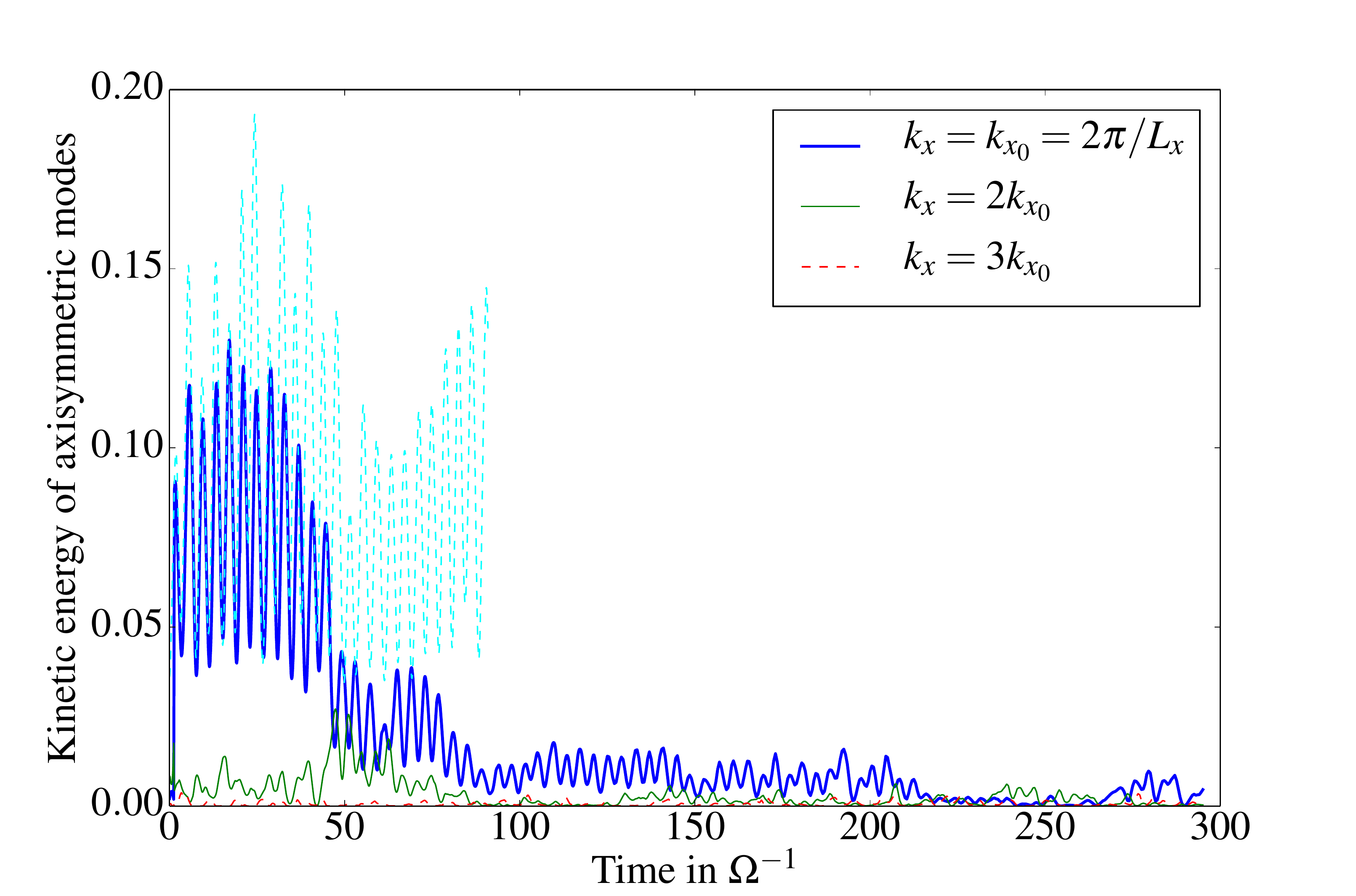}
 \caption{Time-evolution of the fundamental axisymmetric mode $k_x=k_{x_0}$ for the MHD (SGMRI-20, blue curve) and for the hydrodynamic simulation (SG-hydro, cyan curve). We superimposed in green and red the harmonic modes $k_x=2k_{x_0}$ and $k_x=3k_{x_0}$ for SGMRI-20. Note that for SG-hydro, the axisymmetric mode continues until $t=200\,\Omega^{-1}$ (not show here due to a crude sampling)}
\label{fig_axiwave}
 \end{figure}

One interesting question is the dependence of our results on
initial conditions. Instead of starting from an MRI-turbulent state,
one could imagine starting from a hydrodynamic GI turbulent state, in
which a seed magnetic field is introduced. Will the final state look
like the one described in \ref{starting_MRI}, or will it be different
and thus indicative of hysteresis? To answer this, we
prepared a 3D hydrodynamic gravito-turbulent state (without magnetic
field) with
 $\tau_c=20\, \Omega^{-1}$. This state is obtained from the simulation ``SG-hydro", already described in \citet{riols17b}. At $t=40\,\Omega^{-1}$, we introduced a zero-net-flux toroidal seed field, with sinusoidal shape in $z$ and initial amplitude $B_{y_0}=10^{-3}$.

Figure \ref{fig_average2} shows the evolution of various averaged quantities
computed from the pure hydrodynamic simulation  (blue curve) and
the new magnetic state that it initiates (green curves, labelled 
SGMRI-20). An immediate and important result is the quasi-exponential
amplification of the seed magnetic field by the pre-existing turbulent flow. The amplification lasts for $200\,\Omega^{-1}$ and the dynamo field then saturates at $E_m \approx 0.006$, which is 6000 times larger than its initial value, but still smaller than the average kinetic energy $E_c \approx 0.04$.  Second, we found that the final state is very similar to the one computed in \ref{starting_MRI}, suggesting that it is independent of the initial condition. Note finally that the confinement of the magnetic field is even stronger than for $\tau_c=100\,\Omega^{-1}$ (see Fig.~\ref{fig_byzt} and Fig.~\ref{fig_vfluct})

\subsubsection{Comparison with the pure hydrodynamic GI state}
\label{comparison_hydro}

We next compare the magnetized state with the pure hydrodynamic GI state,
computed for the same $\tau_c=20\,\Omega^{-1}$ (green vs blue curves
in Fig.~\ref{fig_average2}). Magnetic fields do not seem to have a
substantial effect on the thermodynamics and gravitational quantities,
since $Q$, $E_T$, and $G_{xy}$ are much the same. However, there is a non-negligible drop in
kinetic energy and Reynolds stress (roughly a factor 2), indicating
the propensity of the Lorentz force to impede GI motions. Figure
\ref{fig_vfluct} (third and fourth panels) shows that the r.m.s
velocity fluctuations are smaller than in the hydrodynamical run
SG-hydro.  The mean magnetic pressure is far too weak, compared to the
thermal pressure, to compete with self-gravity and interfere with
the linear response of the GI modes. It is more likely that the local
gradients of magnetic fields nonlinearly affect the dynamics of the turbulent waves through magnetic tension. 

To carry out a more in-depth investigation, we analysed the spectra
and in particular the small-scale activity. Recently, the hydrodynamic
GI was found to be subject to
a small-scale parametric instability, probably due to the resonance
between inertial waves and a large-scale epicyclic mode
\citep{riols17b}. The small-scale structures instigated by this
instability  are visible in the density plot on the top right panel of
Fig.~\ref{fig_SGmri_rho}. They take the form of small ribbon-like
fluctuations that disturb the spiral wave fronts. How does
this instability behave in presence of a tiny but non-negligible
magnetic field? Visually, Fig.~\ref{fig_SGmri_rho} (top, central
panel) shows that structures of scale $ \lesssim H_0$ are still
present, in particular in the left part, but they are less
pronounced.  In particular, in the right part where the magnetic field
is stronger, GI spiral waves are almost entirely free of the small-scale
parasitic turbulence, suggesting that magnetic fields suppress the
parametric instability. 

\begin{figure}
\centering
\includegraphics[width=\columnwidth,trim=0cm 5.cm 0cm 0cm, clip=true]{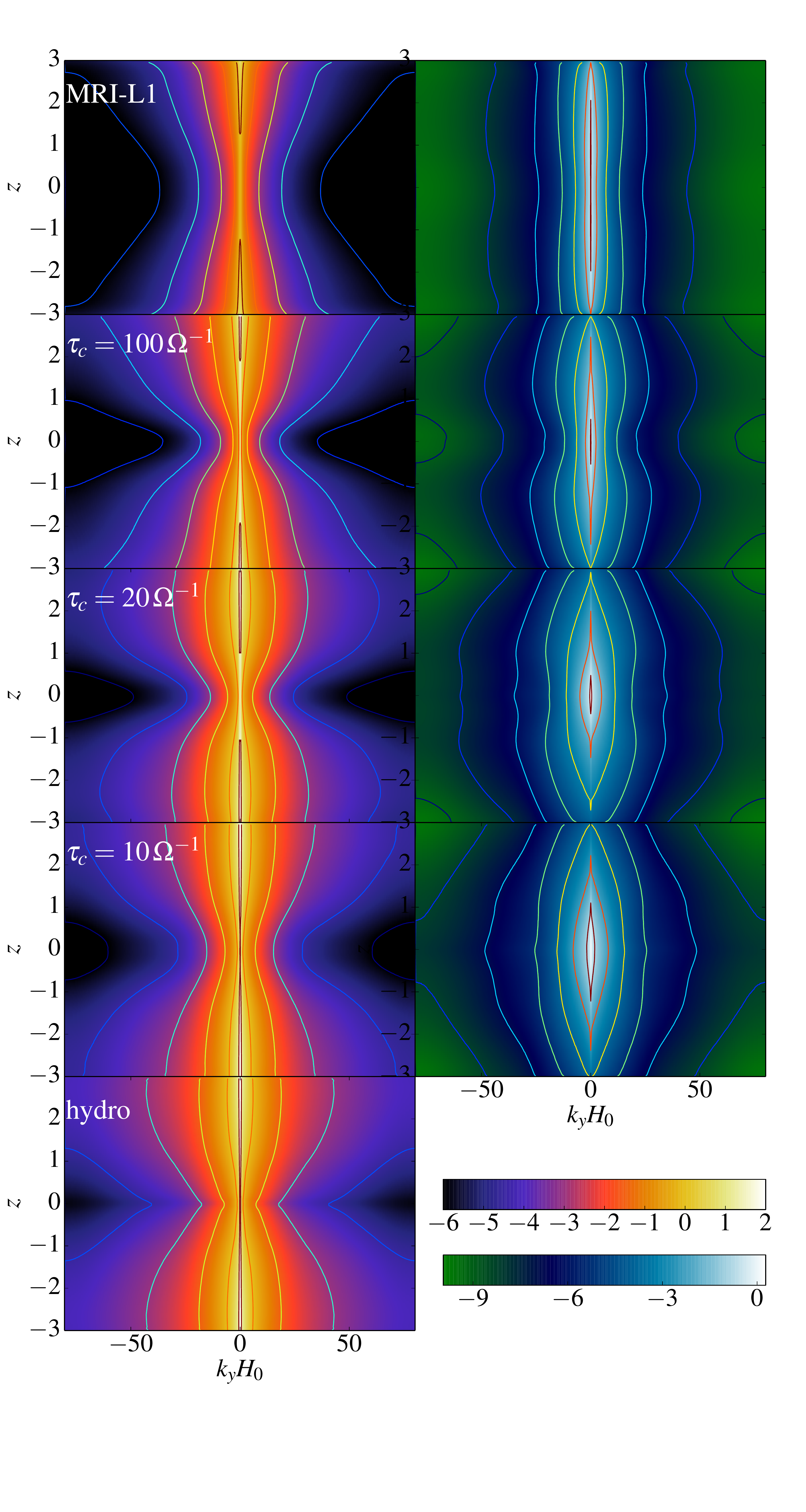}
 \caption{1D kinetic $\mathcal{E}_K(k_y,z)$ (left) and magnetic
   $\mathcal{E}_M(k_y,z)$ (right) power spectrum  as a function of
   $k_y$ and altitude $z$. From top to bottom: 1: pure MRI (MRI-L1),
   2: GI with MHD and
  $\tau_c=100\Omega^{-1}$ (MRISG-100), 3: GI with MHD and
  $\tau_c=20\,\Omega^{-1}$ (SGMRI-20), 
4: GI with MHD and $\tau_c=10\,\Omega^{-1}$ (SGMRI-20) and 5: hydrodynamical GI (SG-hydro), $\tau_c=20\,\Omega^{-1}$.}
\label{fig_spec2}
 \end{figure}

This result can be checked
quantitatively and statistically by plotting the time-averaged
spectrum $\mathcal{E}_K(k_y,z)$ of both simulations, SG-hydro and SGMRI-20. 
 The result, shown in Fig.~\ref{fig_spec2}, clearly indicates that in
 the MHD case
less kinetic energy is found on small scales (large $k_y$). 
Additionally, we found that:
\begin{equation}
\Lambda_{K}^{10,1}(H_0)_{\text{MHD}} 
\simeq  0.17\, \Lambda_{K}^{10,1}(H_0)_{\text{hydro}},
\end{equation}
proving that the ratio of large-to-small scales differ
by about an order of magnitude in the two simulations.

The small-scale parametric modes are thought to be excited by a large
scale axisymmetric oscillation with $k_x=k_{x_0}=2\pi/L_x$
\citep{riols17b}. In
 hydrodynamic simulations 
this mode possesses a large amplitude $\simeq 0.5\, c_s$ and undergoes
regular oscillations at a frequency close to $\Omega$
\citep{riols17b}. We now check what happens to this mode when a magnetic field is
included. Fig.~\ref{fig_axiwave} shows 
the time-evolution of its kinetic energy in the MHD simulation (blue
curve) and in the hydrodynamic simulation (cyan/light curve). 
Strikingly magnetic fields damp and ultimately
kill the large-scale axisymmetric oscillation $k_x=k_{x_0}$. Figure
\ref{fig_axiwave} also shows that the harmonic modes (in particular
$k_x=2k_{x_0}$ and $k_x=3k_{x_0}$)  remain weak and are unimportant. 
By projecting the forces onto the mode $k_x=k_{x_0}$,  we found that magnetic tension and pressure have no direct effect on it. Instead, the field is degrading the nonlinear couplings that feed the axisymmetric mode. 

\subsection{Regime of efficient cooling and fragmentation ($\tau_c \leq 10 \Omega^{-1})$}
\label{efficient_cooling}
\begin{figure}
\centering
\includegraphics[width=0.95\columnwidth]{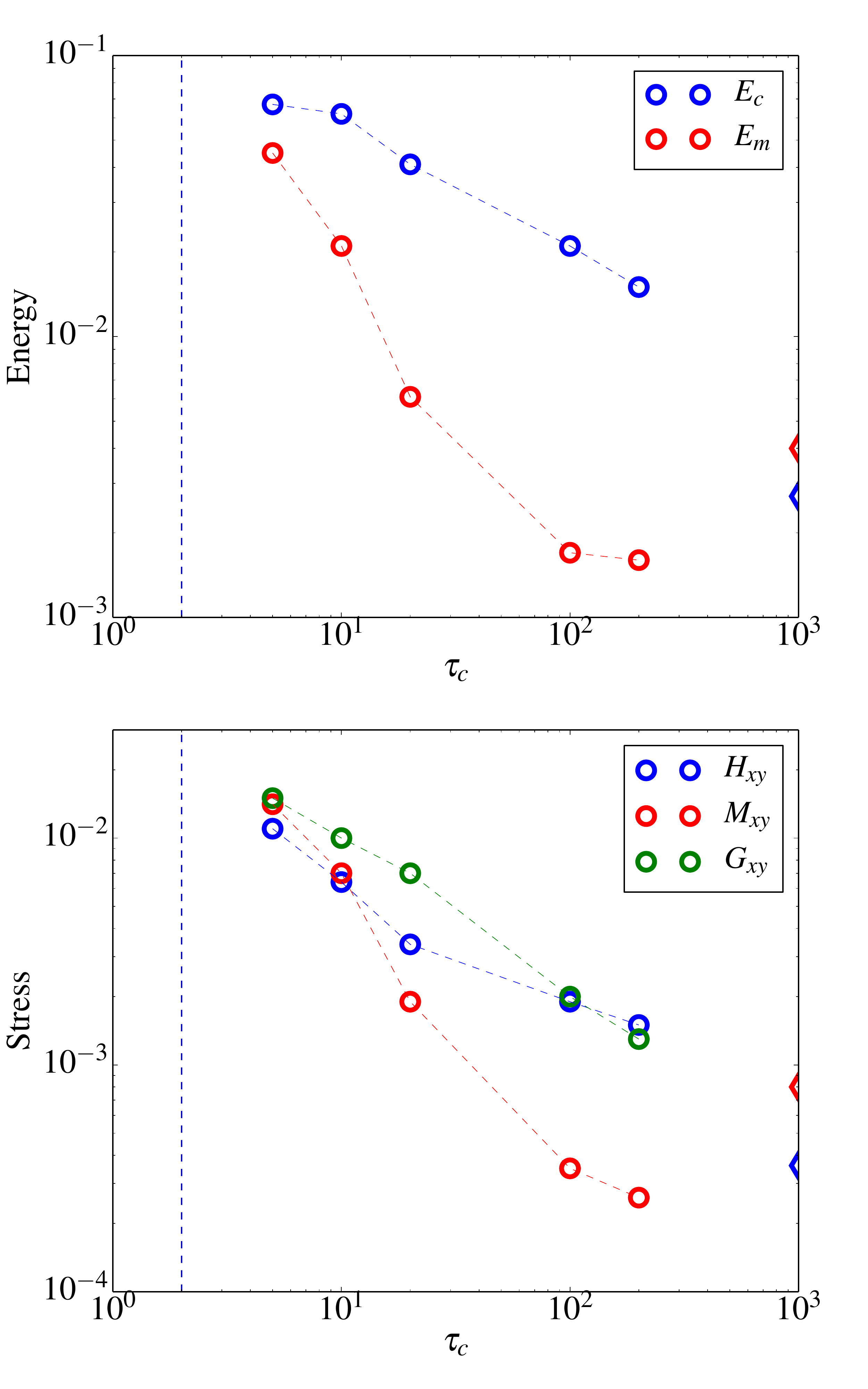}
 \caption{Top: time and box-averaged kinetic and magnetic energy as functions of cooling time $\tau_c$. Bottom: time and box-averaged Reynolds, Maxwell and gravitational stress as functions of cooling time $\tau_c$. The diamond markers on the right margin show these quantities for the pure MRI case.}
\label{fig_eratio}
 \end{figure}
 
The last regime investigated is the one of short cooling
times $\tau_c \leq 10 \, \Omega^{-1}$. Three cases were
considered: $\Omega\tau_c=10$, 5, and 2. Each simulation was
initialized from the neighbour state with longer $\tau_c$. In the
first two cases, $\tau_c=10\,\Omega^{-1}$ and $\tau_c=5\,\Omega^{-1}$,
we simulated gravito-turbulent states for $\sim 100 \, \Omega^{-1} \gg
\tau_c$. Table \ref{table1} shows the averaged quantities
corresponding to these states.  

To compare with simulations of longer
$\tau_c$, we plot in Fig.~\ref{fig_eratio} the mean kinetic and
magnetic energies as well as the different stresses as a function of
cooling time $\tau_c$. This figure displays one of the most
important results of our paper. As $\tau_c$ decreases,
and the disc enters the efficient cooling regime,
magnetic and kinetic energy tend to equipartition. In addition, 
the Maxwell stress
grows larger than the Reynolds stress and attains values comparable to the
gravitational stress.
In this regime, we found that a strong primarily toroidal field dominates,
its morphology similar to the large scale spiral wakes. This is
certainly suggestive that a powerful dynamo
is supported by the spiral waves (see Section \ref{sec_dynamo} for
further discussion).  We checked that the flux  
is again confined near the midplane, between $-H_0$ and $H_0$,
although the confinement appears no stronger than for $\tau_c=20\, \Omega^{-1}$. 

Figure \ref{fig_spec2} (right column) shows the magnetic spectrum for
different $\tau_c$. In addition to the strong large-scale field there
is evidence here of energy on shorter scales when $\tau_c$ 
tends to small values.
These small-scale structures are much more developed than in the pure
MRI case, and may be telling us that reconnection is an important
ingredient sustaining this state.
Indeed, in the low $\tau_c$ regime, 
the flows generate transient plasmoids embedded within magnetic pressure rings (or
magnetic islands) and which we associate 
with reconnection sheets. See the bottom panel of
Fig.~\ref{fig_SGmri_rho} for a 3D rendering of an example plasmoid. The
plasmoids
resemble those appearing in
\citet{riols16a} but are somewhat more marginal and survive for
only a few $\Omega^{-1}$. 

Finally, for $\tau_c=2\,\Omega^{-1}$ the magnetic field becomes even
more intense during the first 15 $\Omega^{-1}$ but the disc fragments
into several bound objects with densities exceeding 
1000 times the background density. The time step becomes of order
$10^{-6}$ 
and makes the study of this state impossible.

In conclusion, for cooling times  $\tau_c \lesssim10 \,\Omega^{-1}$, a
powerful dynamo mechanism amplifies a magnetic field 
to equipartition levels much stronger than what the MRI is capable of.
The field displays both strong large-scale toroidal features and
 small-scale non-axisymmetric structure.  
Fragmentation occurs for cooling times very similar to those obtained
in 3D hydrodynamic GI \citep{shi2014,riols17b}. Regarding this last
point, we warn the reader that the fragmentation criterion probably
depends on
numerical resolution. Transient clumps are not much larger than the
grid size and so we expect numerical diffusivity to weaken magnetic
tension on these scales. Magnetic tension might otherwise facilitate
gravitational collapse because of its breaking of angular momentum
conservation.
 
\subsection{Effect of a strong imposed $B_y$ or a net $B_z$}
\label{field_geometry}
To study the dependence of our results on magnetic field geometry,  we
performed two simulations, one with a strong initial imposed toroidal
field, the other with a net vertical field.  In both
cases $\tau_c=20\Omega^{-1}$. 

In the first run, labelled
SGMRI-20-By0.1, the net toroidal flux in the box, initially $\langle
B_y\rangle=0.1 $, is free to evolve during the simulation. 
Our aim is to make a comparison with 2D simulations
\citep{riols16a}, in which a mean toroidal field was imposed. Table
\ref{table1} shows that the magnetic perturbations reach almost
equipartition, however the average temperature and $Q$ in the box 
fail to rise significantly. In contrast, $Q$ was found to be 3
times larger than its hydrodynamic value in 2D. The main reason for
this difference is that in 3D internal
energy is released via vertical outflows, and this always prevents
the disc from heating up inordinately. It is however not excluded that this result depends on the vertical extent of the box. Reconnection sheets are also less active
than in 2D and plasmoids are only marginally produced. A ``hotter" and more active state could possibly emerge with a  larger initial toroidal flux.

In the second run, labelled SGMRI-20-Bz0.1, the net vertical flux
$\langle B_z\rangle=0.1 $ is conserved during the simulation. This
corresponds to an initial midplane plasma $\beta = 2\Sigma_0 H_0 c_{s_0}^2/(\gamma
B_z^2) \simeq 225$.  A strong magnetic field builds up in a short
period of time $\simeq 10\, \Omega^{-1}$. The final state consists of
a magnetically-dominated disc with $E_m \gtrsim E_c$ and $M_{xy}\gg
G_{xy}$. The gravitational stress is substantially smaller than in the
zero-net-flux case. The transport efficiency is very high ($\alpha
\simeq 0.54$) and reminiscent of the MRI state obtained by
\citet{salvesen16} in the limit of small beta plasma and $Q\rightarrow
\infty$. To sustain such states thermodynamically, the outflows need
to be quite powerful, so that $\tau_w \gg \tau_c$. 

By checking the
density structures, we found that spiral GI waves are completely
dominated by the MHD dynamics; unlike the zero-net flux case, the MRI
seems to be strong enough to inhibit the GI.  This result is
nevertheless very preliminary. A more detailed investigation, spanning
different  $\langle B_z\rangle$ will be presented in the future.  
The main point to take away is that the dominance of GI over MRI,
witnessed in Sections 4.3 and 4.4, is not inevitable and that there exists
at least one regime where the reverse situation holds. 

\begin{figure}
\includegraphics[width=\columnwidth]{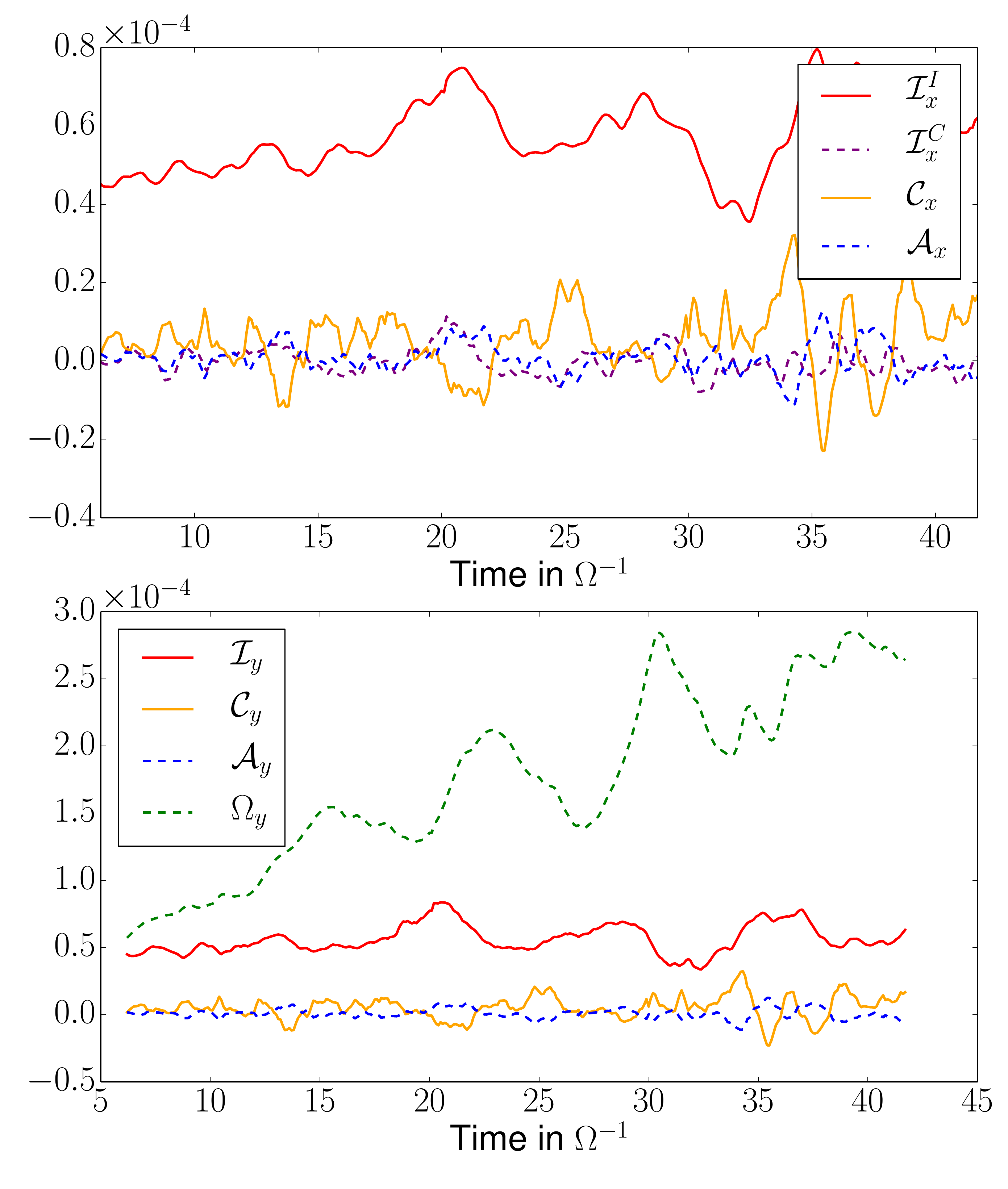}
\caption{Evolution of the magnetic energy budget for $B_x$ (top) and $B_y$ (bottom) during the magnetic growth phase in SGMRI-20 ($\tau_c=20\Omega^{-1}$). Each curve corresponds to a term in Eq.~\eqref{eq_magenergy}, averaged in space between $z=-1.5 H_0$ and $z=1.5 H_0$. Note that numerical dissipation of magnetic energy contributes to the budget (negatively), but is not represented here.}
\label{fig_budget}
\end{figure}
\begin{figure*}
\includegraphics[width=1.01\textwidth]{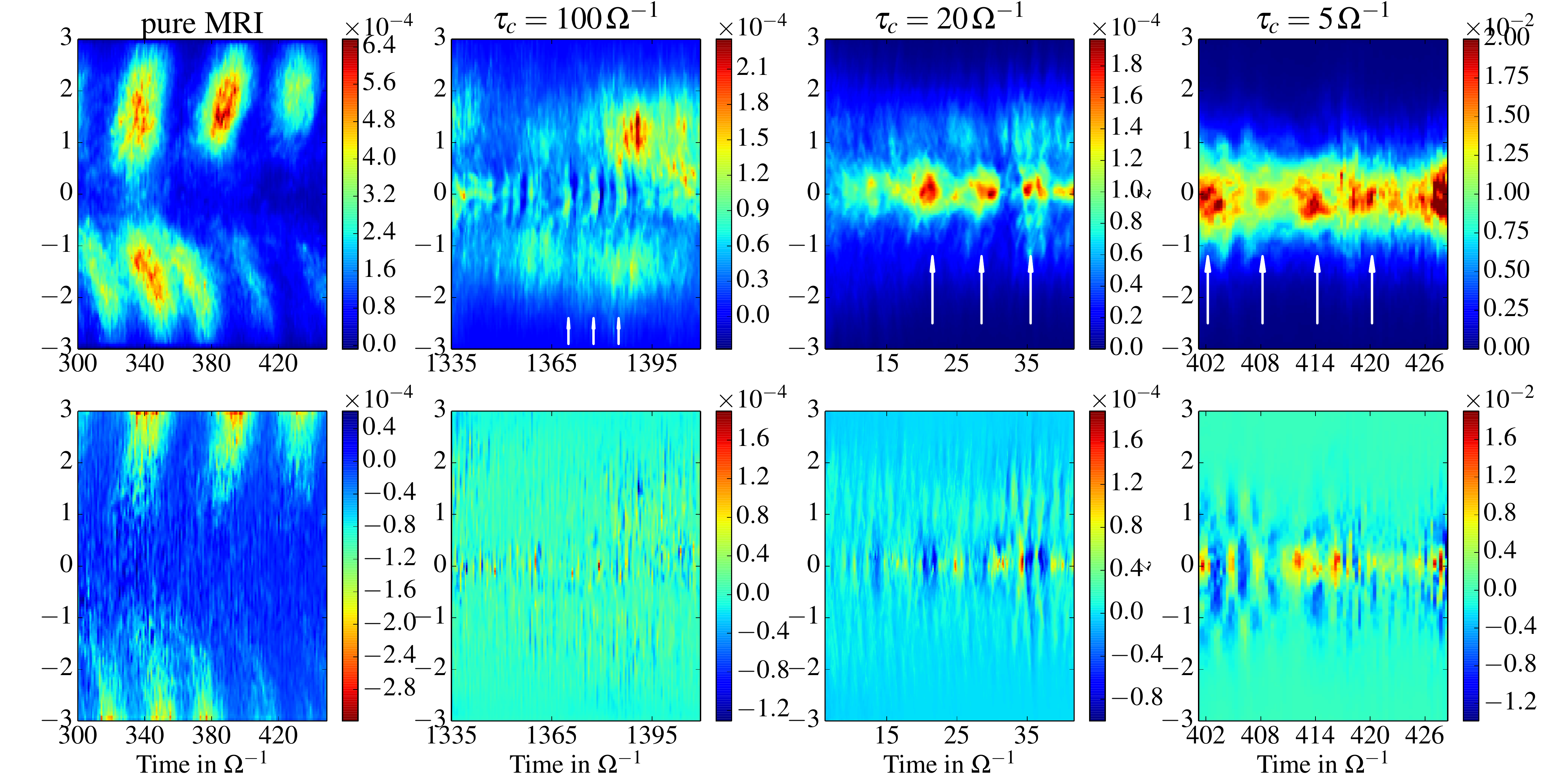}
\caption{Space-time diagram $(t,z)$ showing the sources of magnetic
  energy in MHD gravito-turbulence, $\mathcal{I}_{x}^I$ (top panels)
  and $\mathcal{C}_{x}$ (bottom panels). From right to left: $\tau_c=5$, $\tau_c=20$, $\tau_c=100\,\Omega^{-1}$ and pure MRI turbulence ($\tau_c=\infty$). Each diagram is averaged in $x$ and $y$. White arrows show the extrema of the dynamo cycle where $\mathcal{I}_{x}^I$ is maximum. They are regularly spaced by intervals of $6 \, \Omega^{-1}$ for the last plot ($\tau_c=5$) and  $7\, \Omega^{-1}$ for the second and third plots ($\tau_c=20 $ and $\tau_c=100 $ ). Note than in the pure MRI case (leftmost panels), the interval of time is much wider than in other case, as the dynamo cycle lasts longer ($\sim 50\, \Omega^{-1}$ for one reversal, i.e $\sim 16$ orbits for the full cycle). }
\label{fig_bfield}
\end{figure*}

\section{From MRI to spiral wave dynamos}
\label{sec_dynamo}

The aim of this section is to characterise in more detail the dynamo process
responsible for the large and small scale magnetic field in 3D
gravito-turbulence. The results of Section
\ref{comparison_hydro} indicate clearly that the parametric instability is
suppressed by a magnetic field, at least in cases $\tau_c\leq
20\,\Omega^{-1}$, and thus we exclude a dynamo supported by helical inertial waves in what follows. In Section \ref{MRIGIsim}, we showed that the MRI
is partially impeded by the GI motions at $\tau_c\sim 100\Omega^{-1}$, 
and probably eliminated at shorter $\tau_c$ (Sections \ref{intermediate_cooling} and  \ref{efficient_cooling}). 
It is imperative then to understand at lower $\tau_c$
 whether magnetic activity is exclusively  
sustained by GI motions or if the MRI, though weak, remains important. 
If the MRI is negligible in that regime, then we must also 
reveal when and how the transition between
the MRI and spiral wave dynamos takes place.  

\subsection{Magnetic energy budget and Helmholtz decomposition}
\label{magnetic_budget}

To aid our understanding of the magnetic field generation we introduce
certain diagnostics derived from the magnetic energy budget. 
The evolution of the box averaged magnetic energy for each component $B_i$, where $i={x,y,z}$, is:
\begin{equation}
\dfrac{1}{2} \dfrac{\partial \langle B_i^2 \rangle}{\partial t}= \mathcal{I}_i+\mathcal{A}_i+\mathcal{C}_i+\mathcal{D}_i+{\Omega}_i
\label{eq_magenergy}
\end{equation}
where
\begin{equation}
\label{eq_induc_terms}
\mathcal{I}_i=B_i B_j \dfrac{\partial u_i}{\partial x^j} \quad  \mathcal{A}_i=- \dfrac{1}{2} u_j \dfrac{\partial B_i^2}{\partial x^j} \quad \mathcal{C}_i=- B_i^2 \,\left(\nabla\cdot \mathbf{u}\right) \\
\end{equation}
and $\Omega_i=-S B_xB_y\delta_{iy}$. The summation on the index $j$ is
implicit in (\ref{eq_induc_terms}). The first term $\mathcal{I}_i$ is
the induction or ``stretching" term, the second  $\mathcal{A}_i$
corresponds to advection of the field and the third  one
$\mathcal{C}_i$ is associated with the compression or expansion of the
flow,  allowing conservation of magnetic flux in compressible
fluids. Finally $\mathcal{D}_i$ denotes all form of dissipation
(numerical or Ohmic) and  $\Omega_i$ is the $\Omega$-effect (linear
stretching by the shear). 

The stretching term $\mathcal{I}_i$ can be
decomposed into two parts, one related to incompressible
fluid motions, the other related to compressible motions. 
To distinguish the two, we use the Helmholtz decomposition:
\begin{equation}
\label{eq_helmholtz}
\mathbf{u}=\mathbf{u}_{\text{c}}+\mathbf{u}_{\text{ic}}=\mathbf{\nabla}\varphi + \mathbf{\nabla}\times\mathbf{{\Psi}},
\end{equation}
where $\varphi$ is a scalar field, defined up to a constant and $\mathbf{\Psi}$ a vector field defined up to a gradient field. The
first term is the compressible part of $\mathbf{u}$ and is curl free.  The second term is the
incompressible or solenoidal part, and it 
is divergence free. The details of their calculation are explained in \citet{riols17b}. We can then write:
\begin{equation}
\mathcal{I}_i= \mathcal{I}_i^C+\mathcal{I}_i^I=B_i B_j \dfrac{\partial (\nabla \varphi)_i }{\partial x^j}+B_i B_j \dfrac{\partial (\nabla \times \mathbf{\Psi})_i }{\partial x^j},
\label{eq_induction_term}
\end{equation}
where the superscripts $C$ and $I$ indicate compressible and
incompressible respectively.

Figure \ref{fig_budget} shows the energy budget associated with the
generation of $B_x$
and $B_y$ in SGMRI-20 averaged over the box at altitudes $z<1.5 H_0$. The
calculations were performed during the early phase of the simulation 
when the magnetic field
is amplified (between $t=0$ and $t=50\Omega^{-1}$), but we confirmed that
the budget is similar in the saturated state. It is clear that the
radial field grows primarily via the incompressible stretching
term $\mathcal{I}_x^I$ (i.e driven by solenoidal motions). Advection
and compressible effects are negligible.
On the other hand, the toroidal field is generated primarily via the
$\Omega$-effect, although $\mathcal{I}_y^I$ also participates in the dynamo
action to a lesser extent.

 We found the dominance of the same dynamo
terms for all $\tau_c$ that we tried (100, 20, and 5). Even at
$\tau_c=5$ the compressible contribution $C_x$ is at most 10\% the
solenoidal contribution. It is important to note, however, that the
spiral density waves consist of both incompressible (vortical) and
compressible motion (e.g.\ Figs 12 and 13 in Riols et al.~2017). 
 Because the maintenance of $B_x$ is
critical for the dynamo, we focus exclusively on that component in what
follows.

\subsection{A transition between two dynamo processes}
\label{transition_dynamo}

We take as a starting assumption that the system can sustain two different
dynamos: one driven by GI spiral motions at short
cooling times ($\tau_c \leq 20\, \Omega^{-1}$) and another driven by
the MRI at long cooling times ($\tau_c\gg 100\, \Omega^{-1}$). In the
intermediate regime between these cooling limits things are less
clear. 
In this section we concentrate on this `mixed' regime to
determine the transition between the two dynamos.

Figure \ref{fig_bfield} displays the horizontally-averaged
induction term $\mathcal{I}_x^I$ (top panels) and compression term
$\mathcal{C}_x$ (bottom panels)
as a function of time and altitude
$z$, for different states and cooling times. The first (extreme) case
corresponds to the pure MRI simulation without self-gravity (far
left column). Magnetic energy is produced above the midplane, at $z\gtrsim
H_0$ and at regular intervals of time. The evolution of
$\mathcal{I}_x^I$, especially, resembles the butterfly diagram in
Fig.~\ref{fig_byzt}. The period is slightly shorter than 50
$\Omega^{-1}$ (i.e 8 orbits) which corresponds to 16 orbits for the
full MRI dynamo cycle (we are plotting energy quantities here and not
field polarity). Within the butterfly wings, $\mathcal{I}_x^I$
fluctuates on short timescales $\simeq \Omega^{-1}$ comparable to the
MRI growth time. The compressible component
$\mathcal{C}_x$, on the other hand, 
is negligible in the disc and even negative in the corona, 
indicating that magnetic flux is relaxed through the acceleration of the flow induced by MHD winds.

The diametrically opposite case corresponds 
to MHD gravito-turbulence with very short cooling, $\tau_c=5 \,
\Omega^{-1}$ (far right column). Here 
the magnetic field is generated exclusively in the midplane regions
where GI motions are stronger. Very little is occurring at $z\simeq
H_0$ and the `fluttering' of $\mathcal{I}_x^I$ at
$\sim  \Omega^{-1}$ is less pronounced. Instead the magnetic field generation
oscillates on longer periods of $\sim 6-7\, \Omega^{-1}$ (i.e
an orbit)  which correspond to the typical lifetime of a large-scale
density wave. In addition, the compressible component
$\mathcal{C}_x$ is larger and can take values near $\mathcal{I}_x^I$ in the
midplane, although it is cancelled out, on average, by
breathing motions in the corona.  There is no
MRI dynamo `signature' here, as least as it appears in the leftmost columns. 
These 
results suggest that at $\tau_c=5 \, \Omega^{-1}$ magnetic field
generation, and hence the dynamo, is
categorically different to that associated with the MRI
and  controlled entirely by the GI spiral motions.

What about the intermediate regime? At $\tau_c=20$, $\mathcal{I}_x^I$
is localised to the midplane, a feature we associate with the GI. 
And indeed, a distinct cycle of period 7 $\Omega^{-1}$ ($\sim$ 1 orbit) emerges as soon 
as the magnetic field starts to grow (indicated by the white arrows). 
However, some faint activity can be detected at $z\simeq H_0$,
which we attribute to a very weak remnant MRI. (The lower altitude of the MRI
signature is due to the vertical compression of the disc when
self-gravity is included.)   

In the case of $\tau_c=100\Omega^{-1}$, we can distinguish two magnetic
sources of similar magnitude. The first is localized in the corona
above $z=H_0$ and fluctuates rapidly (see for instance the red spots
at $t=1380\,\Omega^{-1}$ or the thin plumes on the left side at
$t=1335\, \Omega^{-1}$). These disorganized and intermittent
structures we interpret as a weak MRI, attempting to
survive in the turbulent GI background. In the next subsection we show
that
these features are suppressed by resistivity, 
which reinforces our belief that they are
MRI-related. The second source is
localized at the midplane and again composed of coherent and
regular structures of period $\simeq 7\, \Omega^{-1}$. These we
associate with the GI. These
two magnetic sources seem to interfere in a destructive way, probably
explaining why the magnetic activity is very inefficient in the mixed regime.\\
\begin{figure}
\centering
\includegraphics[width=\columnwidth]{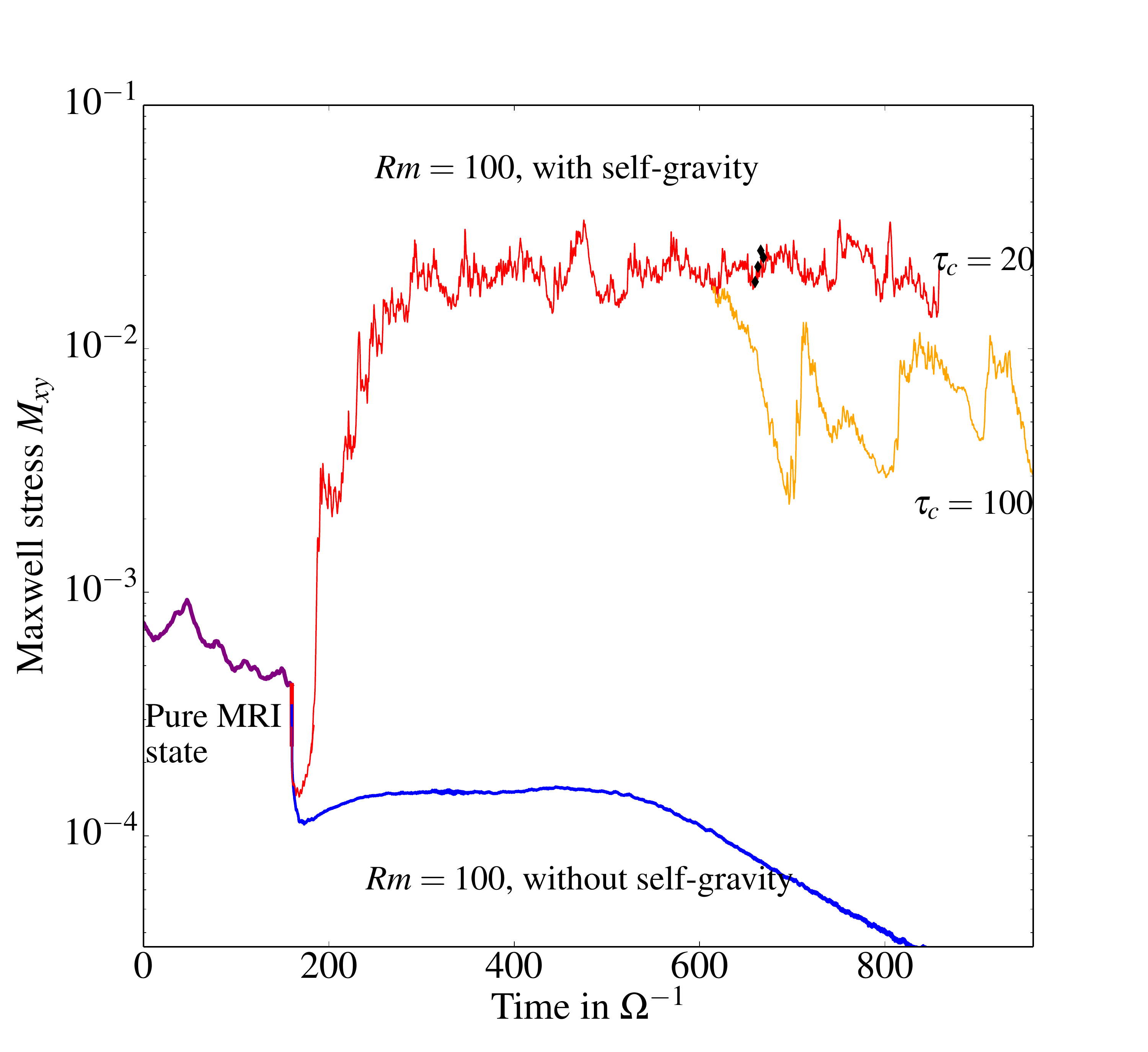}
 \caption{Time-evolution of the Maxwell stress for different cases. The purple curve is our initial condition, taken from the pure MRI simulation (MRI-L1), the blue curve corresponds to the same setup but with $\text{Rm}=100$. Finally the red ($\tau_c=20\,\Omega^{-1}$) and orange curves ($\tau_c=100\,\Omega^{-1}$) correspond to the case with self-gravity and  $\text{Rm}=100$ }
\label{fig_resistive}
 \end{figure}

\subsection{Dynamo action in highly resistive flows}

One may ask if these two dynamo processes
survive in PP discs where non-ideal effects, such as Ohmic
diffusion, can be important. It is known from numerical simulations
that the MRI dynamo breaks down for $R_m=\Omega H_0^2/\eta$ less than
a few hundred in the most favourable configuration
\citep{oishi11,riols15}. Is it still possible to maintain a
pure GI dynamo for such Rm?  
Are magnetic fields still generated  in the 
intermediate/large cooling regime if the MRI is suppressed?
To answer these questions, we performed several simulations with
explicit Ohmic resistivity. The magnetic Reynolds number is fixed to
100 in order to quench the MRI completely.  

First, we started from a
state drawn from simulation MRI-L1, added magnetic diffusivity but no
self-gravity (and kept $\tau_c=\infty$). As expected,
Fig.~\ref{fig_resistive} (blue curve) shows that the Maxwell stress
tends to 0 and all activity dies away. Next we redid the simulation but
with magnetic diffusivity and self-gravity (and set
$\tau_c=20\Omega^{-1}$), the result
being the red curve. As is clear, magnetic activity
is sustained for long times and tends toward a
quasi-equilibrium state. We found that this state is highly magnetized
($M_{xy} \gg G_{xy}$), despite the strong Ohmic resistivity and the
absence of any MRI. This result differs to the 2D case
where Ohmic diffusion \emph{reduced} the magnetic to kinetic
energy ratio \citep{riols16a}. This is a strong clue that dynamo
action (precluded in 2D) is responsible for the strong fields
witnessed here. 

We also performed a simulation with $\tau_c=100$ starting from a quasi-equilibrium state at
$\tau_c=20$. The dynamo is not suppressed, although the magnetic activity is reduced
by a factor 5.  We checked that the dynamo operates
exclusively in the midplane, indicating that the component at $z=H_0$,
obtained without resistivity, was indeed the manifestation of the MRI.   
Note that the resolution we used was $128\times 128 \times 64$, but we checked
that similar states seems to hold for at least $30 \,\Omega^{-1}$ 
at higher resolution ($512\times 512 \times 128$). 

In conclusion, the GI dynamo is a robust mechanism that works
independently of the MRI. It is also able to amplify magnetic field to
equipartition even in regimes of relatively strong Ohmic
resistivity. Potential implications for PP discs and their  
dead zones are discussed in Section \ref{sec_discussion}.
\begin{figure*}
\includegraphics[width=\textwidth]{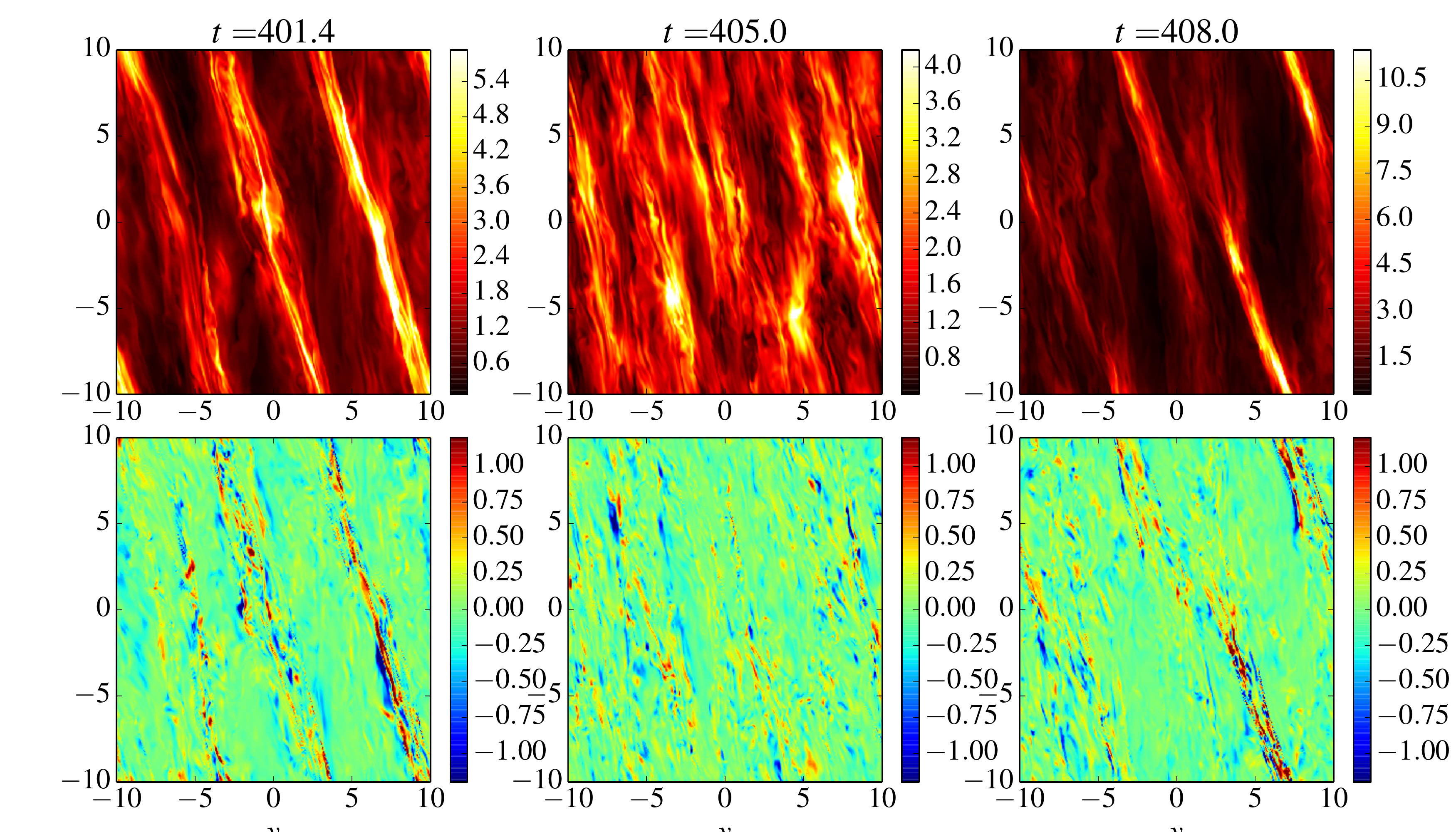}
\caption{Snapshots of density (top) and $\mathcal{I}_x^I$ (bottom)
  from the same simulation with $\tau_c=5\Omega^{-1}$ taken at 3
  different times. 
The first and third columns ($t=401.4\,\Omega^{-1}$ and
$t=408\,\Omega^{-1}$) 
correspond to a maximum in the dynamo cycle, 
while the central column ($t=405\,\Omega^{-1}$) corresponds to a minimum.}
\label{fig_dynamospiral}
\end{figure*}

\subsection{The spiral wave dynamo} 
\label{GIdynamo_spirals}

Our simulations show that
the GI dynamo, distinct from the well known MRI dynamo, 
is very efficient at generating and sustaining both large and small
scale magnetic fields (see in
particular Fig.~\ref{fig_spec2}). Our next task is to
determine how the dynamo process works. Though we leave a detailed
exploration to a future publication, we outline in this section
some of its fundamental features.
In particular, we discuss the nature of the cycle observed in 
Section \ref{transition_dynamo} and clarify the role of the spiral waves. 

Figure \ref{fig_dynamospiral} shows three planar snapshots of the
density and induction
term $\mathcal{I}_x^I$ for a cooling time of $\tau_c=5\,\Omega^{-1}$. 
The three snapshots are at times that correspond to maxima of the cycle (left and
right column) and the minimum (central column). In fact, the $t=401.4$ and $t=408$
snapshots correspond to the two first arrows in the top right panel of
Figure \ref{fig_bfield}. By comparing the morphology of $\rho$ and
$\mathcal{I}_x^I$, 
these figures reveal that
dynamo action is enhanced
when the density spiral waves are strong and well-developed (first and
last columns). Magnetic
energy is produced within the high density regions of the wave, where
compression is maximum. Though the spiral waves have large-scale
structure, magnetic field nonetheless is stretched into
thin
filaments, strongly correlated with the current $(\nabla \times
\mathbf{B})^2$. In contrast, during the minimum of the cycle,
the spiral waves are mixed up and incoherent. The production of radial field
is concomitantly less efficient. It is possible that the destriction
of the spiral waves
results from their interaction with the magnetic field generated
at the earlier stage. We then tentatively 
suggest a cyclic mechanism by which: (a)
spiral waves, amplified by GI, produce local magnetic fields;  (b)
these waves break down into a mixed turbulent state because of magnetic
tension; (c) magnetic fields decay, due to mixing and turbulent
diffusion; (d) new spiral waves grow again. 

In this putative cycle, it is step (a) that is perhaps the least
straightforward. It can be envisaged, however, that a strong density
wave (even axisymmetric) comprises fluid motions that are propitious
for kinematic dynamo action. Its significant radial compression, allied with
baroclinicity, may generate a
vortical flow in the $x-z$ plane, an important ingredient for
small-scale dynamo action  \citep{brand05}. But this poloidal
circulation when combined with the orbital
shear provides helicity, important for large-scale dynamo action. 
In fact, the efficient Ponamarenko and Roberts dynamos bear
similarities to the helical velocity fields a spiral wave may
generate (Ponamarenko 1973, Roberts 1972). If the GI dynamo operates
analogously, then magnetic diffusion is an additional and important
ingredient, either supplied numerically by the grid (in simulations)
or non-ideal MHD (in realistic disks). These concepts will be further
explored and refined in future work.

\section{Conclusions and astrophysical implications}
\label{sec_discussion}

We present a set of 3D shearing box simulations of
magnetised accretion disks that
combine magnetorotational and  gravitational instabilities.  
Our first main result is that both
instabilities coexist only for long cooling times  $\tau_c \gtrsim 100\,
\Omega^{-1}$, with the zero-net-flux MRI suppressed by GI at smaller
$\tau_c$. This result is not necessarily obvious, given that
  the GI modes are large-scale and rather separated from the MRI
  scales. Moreover,  unstratified MRI simulations with generic
  box-scale hydrodynamic forcing appear to strengthen the MRI
  \citep{workman08}. One possibility is that gravitoturbulent
  structures twist, advect, and stretch magnetic field on a timescale
  comparable or shorter than the characteristic MRI time, possibly
  impeding the MRI growth mechanism. Another possibility is that the
  heat produced by GI in the corona increases the
  Brunt-Vaisala frequency leading to suppression of magnetic buoyancy,
  an important ingredient in the MRI dynamo. 
 Finally, we point out resolution problems faced by our numerical 
 set-up. These might also play a role in the MRI's suppression,
 especially if the GI shocks locally increase the numerical diffusivities.\\

Our second main result is that on
shorter cooling times, magnetic fields are amplified
to nearly equipartition strengths via the action of GI density spiral
waves. We explore aspects of this GI (or spiral wave) dynamo, 
in particular its ability to sustain magnetic
fields in highly resistive plasma. Note that the magnetic field
generated by the GI dynamo is significantly greater than that
supported by the MRI.
When a strong net vertical flux is imposed
the system behaves very differently. In
particular, for plasma $\beta \sim 200$, the disc is highly
magnetized, with $\alpha \simeq 0.5$ and the MRI dominates  
over gravito-turbulence.
This third result demonstrates that the MRI's suppression is not
inevitable.\\

It is valuable to compare these results with those obtained by
\citet{fromang04} and \citet{fromang2005} in global disc
simulations. In the case of an initial zero-net-flux toroidal field, they
found that the disc becomes highly magnetized and the MRI weakens the
GI. This state is probably very much determined by the strong field they
imposed initially ($\beta \simeq 8$); in fact, this outcome resembles
the state described in Section \ref{field_geometry}
with an imposed toroidal flux, although our initial field ($\beta \simeq
200$) is much weaker. In the case of a vertical flux, Fromang et al.\ used an initial
$\beta$ of 300, similar to our run SGMRI-20-Bz0.1, 
and found that essentially the MRI impeded GI, in
agreement with our results. It must be stated that a direct comparison is
difficult: the Fromang simulations, in addition to being global, 
 feature a much lower resolution, were run for less 
than 10 orbits (at the inner disc edge), and did not have explicit
cooling (thus allowing the gas to heat up and the $Q$ to increase over
the course of the simulation).\\

A second point of comparison is the razor-thin 2D simulations
of \citet{riols16a} who imposed a toroidal field on the
domain. 
Unlike in 2D, our 3D runs failed to
heat up to large values of $Q$, mainly because vertical outflows carry
a non-negligible amount of heat from the box (especially as the disk vertically
expands). 
And yet, the two models do show
similarities: for instance, the existence of strongly magnetized
states (at low cooling times) characterised by plasmoids and reconnecting current sheet
networks. Like in 2D, the
fragmentation criterion is not strongly affected by the
magnetic field: the critical cooling time is still $\simeq \Omega^{-1}$. This is also in agreement with the 3D
SPH simulations of \citet{forgan17}. In this regime, the MRI is
probably inoperative. However, more work is
required to determine whether this conclusion holds in disc threaded by a
 vertical field. 

We now apply our (admittedly preliminary) results to various
astrophysical systems. First, in the absence of net fields,
we expect no great revision of the
truncation radius for AGN nor the fragmentation criterion in PP disks.
But this conclusion may completely change with a sufficiently
strong net vertical flux (and/or better numerical resolution perhaps). 
In fact, one might envision a disk classification centred
on this distinction, given that the dynamical outcomes are so starkly different. 
While it may be argued that AGN at $>0.01$ pc scales
do not possess large-scale magnetic fields of any great magnitude, this is
unlikely to be the case for class 0 protoplanetary disks, which are
pierced by a relatively ordered field inherited from their natal
clouds and further concentrated by gravitational collapse. Our
zero-net-flux results are perhaps of least relevance to
such disks (which may also be too thick to be well described by
shearing boxes).
Moreover, in the PP disk context the primary
 competition may not be between GI and the MRI as such, but between the GI and
 ambipolar-assisted MHD winds and/or powerful
 Hall-assisted zonal fields (Bai and Stone 2013, Lesur et al.~2014,
 Simon et al.~2015).

Our preliminary study of the GI dynamo (at lowish $\tau_c$) indicates that it is a powerful
and robust way to amplify large and small-scale magnetic fields.
This dynamo has been found to work for magnetic
Reynolds numbers as low as 100.
This could have a number of applications,
especially to the FU-Orionis outburst cycle and more generally to
the dead zones of PP discs where non-ideal MHD is so dominant. 
Indeed, one can imagine that the GI, resulting from the
accumulation of mass in dead zones, could produce its own magnetic
fields well before the ionisation permits the onset of the MRI. A
significant mass could then be accreted in these regions due to
magnetic torques (produced by GI only) leading to `colder' and
probably less violent outburst cycles.

Finally, the GI dynamo may have 
applications in the theory of magnetic field generation in disk
galaxies. 
Current observations suggest
that some spiral galaxies (M81, M51 and possibly M33) are dominated by
bisymmetric-spiral magnetic fields, which have led theorists
to invoke the 
action of spiral waves in their generation. In particular, \citet{chiba90} and \citet{hanasz91} 
proposed a parametric swing excitation due to resonances
between weak magnetic wave oscillations and spiral waves. More
sophisticated models of this parametric resonance have been developed
by \citet{moss97} and \citet{rodhe99}. Most of
these models rely heavily on mean-field theory, for which the microscale
interstellar turbulence, driven by shocks and supernova, is
parametrized by an ``alpha" effect, an absolutely crucial ingredient
for the dynamo
action proposed \citep{elstner00, rudiger04}. In other words,
spiral waves have not been regarded as capable of amplifying magnetic
fields on their own, as our simulations suggest they can.

Our results suffer from several limitations which point towards
a great number of future numerical projects. Our treatment of the
net-vertical-flux case is brief and only examines a single plasma
beta; both stronger and weaker imposed fields should be trialled.
In addition, the numerical robustness of the shearing box in the strong field
case must be properly interrogated (with and without self-gravity). 
We identified some features of the
spiral wave dynamo, but its detailed nature and properties require
further elucidation. In
particular, future work should determine its critical Rm, whether it is slow or fast, 
and whether it is linear (kinematic) or nonlinear. All our runs adopt
a simple Newtonian cooling prescription, and this must be improved
upon in the future. The mean vertical structure exhibited by realistic radiative models
may deviate from our simulations, leading to different vertical motions,
and different (or even no) dynamo behaviour. Finally, non-ideal MHD
must be incorporated piece by piece so as to better mimic the relevant
radii in PP and AGN disks. Does the spiral wave dynamo work when
ambipolar, rather than Ohmic, diffusion holds sway? What is the
influence of the Hall effect? How does GI interact with magnetically
driven winds, and strong zonal fields? As is clear from this
(non-exhaustive) list, there are more than enough
numerical directions and science questions here to keep researchers busy
for quite some time. 

\section*{Acknowledgements}

The authors would like to thank the anonymous reviewer for his comments,
and Ji-Ming Shi, Sebastien Fromang, and Charles Gammie for generously
providing feedback on an earlier version of the paper. This work is
funded by STFC grant ST/L000636/1. Some of the simulations were run on
the DiRAC Complexity system, operated by the University of Leicester
IT Services, which forms part of the STFC DiRAC HPC Facility
(www.dirac.ac.uk). This equipment is funded by BIS National
E-Infrastructure capital grant 
ST/K000373/1 and STFC DiRAC Operations 
grant ST/K0003259/1. DiRAC is part of the UK National E-Infrastructure.

\bibliographystyle{mnras}
\bibliography{refs} 

\label{lastpage}
\end{document}